\documentclass[12pt]{article}

\usepackage{latexsym,amsmath,amssymb,amsthm,amsfonts,graphicx,natbib}
\usepackage{array}
\newcolumntype{P}[1]{>{\centering\arraybackslash}p{#1}}
\usepackage{epsfig}
\usepackage{setspace}
\usepackage{bm,xcolor}
\usepackage{hyperref}
\usepackage{graphicx}
\usepackage{caption}
\usepackage{subcaption}
\usepackage{float}
\usepackage{multirow}

\newcommand*\patchAmsMathEnvironmentForLineno[1]{%
      \expandafter\let\csname old#1\expandafter\endcsname\csname #1\endcsname
      \expandafter\let\csname oldend#1\expandafter\endcsname\csname end#1\endcsname
      \renewenvironment{#1}%
         {\linenomath\csname old#1\endcsname}%
         {\csname oldend#1\endcsname\endlinenomath}}%
    \newcommand*\patchBothAmsMathEnvironmentsForLineno[1]{%
      \patchAmsMathEnvironmentForLineno{#1}%
      \patchAmsMathEnvironmentForLineno{#1*}}%
    \AtBeginDocument{%
    \patchBothAmsMathEnvironmentsForLineno{equation}%
    \patchBothAmsMathEnvironmentsForLineno{align}%
    \patchBothAmsMathEnvironmentsForLineno{flalign}%
    \patchBothAmsMathEnvironmentsForLineno{alignat}%
    \patchBothAmsMathEnvironmentsForLineno{gather}%
    \patchBothAmsMathEnvironmentsForLineno{multline}%
    }
\usepackage[mathlines,displaymath]{lineno}	

\usepackage[textwidth=7in,textheight=9in,centering]{geometry}

\include{def}
\def\dispmuskip{\thinmuskip= 3mu plus 0mu minus 2mu \medmuskip=  4mu plus 2mu minus 2mu \thickmuskip=5mu plus 5mu minus 2mu}
\def\textmuskip{\thinmuskip= 0mu                    \medmuskip=  1mu plus 1mu minus 1mu \thickmuskip=2mu plus 3mu minus 1mu}
\def\beq{\dispmuskip\begin{equation}}    \def\eeq{\end{equation}\textmuskip}
\def\beqn{\dispmuskip\begin{displaymath}}\def\eeqn{\end{displaymath}\textmuskip}
\def\bea{\dispmuskip\begin{eqnarray}}    \def\eea{\end{eqnarray}\textmuskip}
\def\bean{\dispmuskip\begin{eqnarray*}}  \def\eean{\end{eqnarray*}\textmuskip}

\def\paradot#1{\vspace{1.3ex plus 0.7ex minus 0.5ex}\noindent{\bf\boldmath{#1.}}}

\newtheorem{algorithm}{Algorithm}
\newtheorem{remark}{Remark}

\newtheorem{lemma}{Lemma}
\newtheorem{proposition}{Proposition}

\newcommand{\diag}{\text{diag}}

\newcommand{\eps}{\epsilon}
\newcommand{\wh}{\widehat}
\newcommand{\wt}{\widetilde}

\def\SetR{\mathbb{R}}

\def\E{{\mathbb E}}                         
\def\V{{\mathbb V}}

\def\a{\alpha}

\def\d{\rm d}
\def\eps{\epsilon}

\def\t{\theta}

\def\l{\lambda}

\def\N{{\cal N}}

\def\KL{\text{\rm KL}}

\def\tr{\text{\rm tr}}

\def\vech{\text{\rm vech}}

\def\cov{\text{\rm cov}}

\def\LB{\text{\rm LB}}

\def\diag{\text{\rm diag}}

\newcounter{mntcomm}

\theoremstyle{definition} 
\newtheorem{examplex}{Example}[section]
  {\popQED\endexamplex}

\begin{document}
\title{Wasserstein Gaussianization and Efficient Variational Bayes for Robust Bayesian Synthetic Likelihood}
\author{Nhat-Minh Nguyen\thanks{\textit{ARC Centre for Data Analytics for Resources and Environments (DARE), University of Sydney, Australia}}
\and Minh-Ngoc Tran\thanks{\textit{Discipline of Business Analytics, University of Sydney Business School, Australia}}
\and Christopher Drovandi\thanks{\textit{School of Mathematical Sciences, Queensland University of Technology, Australia}} 
\and David Nott\thanks{\textit{Department of Statistics and Data Science, National University of Singapore, Singapore}}
}
\date{\empty}
\maketitle
\begin{abstract}
The Bayesian Synthetic Likelihood (BSL) method is a widely-used tool for likelihood-free Bayesian inference.
This method assumes that some summary statistics are normally distributed,
which can be incorrect in many applications.
We propose a transformation, called the Wasserstein Gaussianization transformation, that uses a discretized Wasserstein gradient flow to approximately transform the distribution of the summary statistics into a Gaussian distribution.
BSL also implicitly requires compatibility between simulated summary statistics under the working model
and the observed summary statistics. A robust BSL variant which achieves this has been developed in the recent literature.
We combine the Wasserstein Gaussianization transformation with robust BSL, and an efficient Variational Bayes procedure for posterior approximation,
to develop a highly efficient and reliable approximate Bayesian inference method for likelihood-free problems.\\

\end{abstract}

\section{Introduction}
In recent years, likelihood-free methodology has emerged as a powerful statistical inference tool in situations where the likelihood function is either intractable or unavailable. This approach requires only the ability to generate data given the model parameter under the working model. Synthetic likelihood, proposed by \cite{wood2010statistical}, is an attractive approach to likelihood-free problems, particularly when the summary statistics for the data approximately follow a Gaussian distribution. Synthetic likelihood has been successfully applied in various fields, including biology, physics and finance among others \citep{wood2010statistical,fasiolo2016comparison,barbu2018two}.

This paper focuses on the Bayesian Synthetic Likelihood (BSL) method \citep{Price:JCGS2018,Ong:STCO2018}, which uses Bayesian inference with the synthetic likelihood approximation. The parametric normal assumption for the summary statistic distribution in synthetic likelihood makes it computationally efficient and simplifies hyperparameter tuning compared to other likelihood-free methods like Approximate Bayesian Computation \citep{sisson2018handbook}. However, violating the normality assumption can lead to biased inference (see the toy example in Sec. \ref{sec: toy example}), and makes the estimation process more challenging \citep{priddle2022efficient}. In related research, \cite{priddle2022efficient} use a linear whitening transformation to decorrelate the summary statistics, and show that using a whitening transformation helps reduce the number of model simulations required in BSL.
\cite{an2020robust} propose a semi-parametric method that assumes flexible marginal distributions for the summary statistics, while describing their dependence using a Gaussian copula.

This paper proposes an alternative method to relax the 
normality assumption in the synthetic likelihood method.
We develop a non-linear transformation that moves the summary statistics to a new set of variables that approximately follow a multivariate normal distribution.
 This transformation produces a sequence of distributions that approximates the discretized Wasserstein gradient flow of the Kullback-Leibler functional on the Wasserstein space. We call this transformation the Wasserstein Gaussianization (WG).
Our WG transformation Gaussianizes the summary statistics, unlike the method of \cite{priddle2022efficient} which only decorrelates them. Unlike 
the method in \cite{an2020robust}, our approach still retains the normality assumption on the transformed summary statistics, inheriting the attractive features of the standard synthetic likelihood method.

 Likelihood-free methods rely on comparing summary statistics generated from a working model with the observed summary statistics. However, if the model is misspecified, this can lead to compatibility issues and unreliable parameter inference. 
 \cite{marin2014relevant} and \cite{Frazier:JCGS2021} have both highlighted this issue. See also \cite{Frazier.et.al:2021}. To address it, \cite{Frazier:JCGS2021} propose an approach called robust Bayesian Synthetic Likelihood (rBSL) that can detect model misspecification and provide useful inferences even in the presence of significant model misspecification. In this paper, we propose coupling the WG transformation with the rBSL approach to improve the robustness, efficiency, and reliability of BSL. We refer to this new method as rBSL-WG.

The rBSL approach, and therefore rBSL-WG, uses a set of auxiliary parameters, whose presence complicates the task of approximating the posterior distribution of the model parameters. \cite{Frazier:JCGS2021} use Markov Chain Monte Carlo, which cannot scale well and can be computationally expensive.
This paper develops an efficient Variational Bayes (VB) method for approximating the posterior distribution in rBSL-WG.
Our VB procedure approximates the posterior distribution of the model parameters directly, rather than approximating the joint posterior distribution of the model parameters and the auxiliary parameters. The former leads to a more accurate inference.
The new VB-rBSL-WG approach provides a highly efficient and reliable Bayesian inference method for likelihood-free problems.
Using a range of examples, we demonstrate that VB-rBSL-WG yields reliable statistical inferences
in situations where the normality assumption on the original summary statistics is not satisfied and/or the model is misspecified.

The rest of the paper is organized as follows.
Section \ref{sec:VB-rBSL} reviews the BSL and rBSL methods, then presents our VB algorithm.
Section \ref{sec: WG for BSL} first provides a brief foundation of the Optimal Transport theory and optimization on the Wasserstein space,
then presents the WG transformation.
The examples are presented in Section \ref{sec: examples}, and Section \ref{sec:conclusion} concludes. The computer code is available at \url{https://github.com/VBayesLab/VBSL-WG}.
Further foundation of the differential calculus on the Wasserstein space and technical proofs can be found in the Appendix.

\section{Efficient Variational Bayes for Robust Bayesian Synthetic Likelihood}\label{sec:VB-rBSL}
\subsection{Bayesian Synthetic Likelihood}
Let $y_\text{obs}$ be the observed data, and $\mathcal M$ the postulated parametric model, with model parameter $\theta$, for explaining $y_\text{obs}$.
Let $p(y_\text{obs}|\theta)$ be the likelihood function under model $\mathcal M$, and $p(\theta)$ the prior. We denote the posterior density by
\[p(\theta|y_\text{obs}) \propto p(\theta)p(y_\text{obs}|\theta).\] 
This work is concerned with the likelihood-free Bayesian inference problem where the likelihood function $p(y_\text{obs}|\theta)$ is intractable but it is possible to generate data from the model.
That is, given any value of the model parameter $\theta$, we can generate data $y=y(\theta)$ from $\mathcal M$.

In the likelihood-free inference literature, it is often desirable to work with a set of lower-dimensional summary statistics $s_\text{obs}$ of $y_\text{obs}$; then, we work with the likelihood $p(s_\text{obs}|\theta)$ instead of $p(y_\text{obs}|\theta)$.
Selection of summary statistics has been thoroughly discussed in the literature; see, e.g, \cite{fearnhead2012constructing,prangle2014semi}.
Often, the likelihood $p(s_\text{obs}|\theta)$ is still intractable and one needs a way to approximate it.
Among many alternatives, the synthetic likelihood method of \cite{wood2010statistical} has received considerable attention.   
This synthetic likelihood method assumes that 
\beq\label{eq: bsl normality}
p(s_\text{obs}|\theta)=\N_d\big(s_\text{obs};\mu(\theta),\Sigma(\theta)\big),
\eeq
where $d$ is the length of $s_\text{obs}$, and the mean $\mu(\theta)$ and variance $\Sigma(\theta)$ are unknown functions of $\theta$.
Even under the normality assumption \eqref{eq: bsl normality},
the synthetic likelihood $p(s_\text{obs}|\theta)$ is still intractable 
because $\mu(\theta)$ and $\Sigma(\theta)$ are unknown.
In most cases, these unknown quantities are estimated based on simulated data.
Let $y_1,...,y_N$ be $N$ datasets generated from model $\mathcal M$ given parameter $\theta$, and let $s_1,...,s_N$ be the corresponding summary statistics.
Write 
\beq\label{eq: mean and variance naive}
\wh\mu(\theta)=\frac1N\sum_{j=1}^N s_j,\;\;\;\wh\Sigma(\theta)=\frac1N\sum_{j=1}^N(s_j-\hat\mu(\theta))(s_j-\hat\mu(\theta))^\top,
\eeq
for the sample mean and sample covariance of these summary statistics.
Then, the synthetic likelihood $p(s_\text{obs}|\theta)$ is replaced by $\overline{p}(s_\text{obs}|\theta)=\int \N_d\big(s_\text{obs};\wh\mu(\theta),\wh\Sigma(\theta)\big)p(y_1,\dots, y_N|\theta)dy_1...dy_N$, and Bayesian inference for $\theta$ is based on the posterior
\beq\label{eq: bsl posterior 1}
\overline\pi(\theta)  \propto p(\theta)\overline{p}(s_\text{obs}|\theta).
\eeq 
This is known as the Bayesian Synthetic Likelihood (BSL) method
\citep{Price:JCGS2018,Frazier:JCGS2021}.
Bayesian computation techniques such as Pseudo-marginal MCMC and Variational Bayes can now be used to approximate $\overline\pi$ \citep{Ong:STCO2018,Price:JCGS2018,Tran:JCGS2017,quiroz2021block}.

Working with the covariance estimate $\wh\Sigma(\theta)$ as in \eqref{eq: mean and variance naive}
can be inefficient.
It is because this estimate might be ill-conditioned, especially in high dimensions, and that one needs to compute the inverse 
$\wh\Sigma(\theta)^{-1}$ in approximating \eqref{eq: bsl posterior 1}.
Computing such a matrix inversion can be computationally expensive and error prone. 
Instead, in this paper, we work with the precision matrix estimate
\beq\label{eq: precision}
\wh P(\theta)=N\Big(\Psi_0+\sum_{j=1}^N(s_j-\hat\mu(\theta))(s_j-\hat\mu(\theta))^\top\Big)^{-1},
\eeq
where $\Psi_0$ is a symmetric and positive definite matrix, e.g., $\Psi_0=\epsilon I$ for some $\epsilon>0$.
Denote $\psi_j=s_j-\hat\mu(\theta)$, and let
\beqn
\Psi_s=\Psi_0+\sum_{j=1}^s\psi_j\psi_j^\top,\;\;s=0,1,...
\eeqn
Working with the precision matrix such as \eqref{eq: precision} has proven useful in various scientific fields; see, e.g., \cite{boyer2023asymptotic,godichon2023natural}.
The following proposition states that $\wh P(\theta)$ is always symmetric, positive definite and can be computed without requiring explicit matrix inversion.
Its proof can be found in Appendix B.
\begin{proposition}\label{lemma}
    For any fixed $\theta$, we have that
    \beq
    \wh P(\theta)=N\Psi_N^{-1}\xrightarrow[N \to + \infty]{a.s} P(\theta)=\Sigma(\theta)^{-1},
    \eeq
and that $A_N:=\Psi_N^{-1}$ is symmetric, positive definite and can be computed recursively as follows
\beq\label{eq:recursive}
\begin{cases}
A_0 = \Psi_0^{-1},\\
A_{s+1} = A_{s}-\big(1+\psi_{s+1}^\top A_{s}\psi_{s+1}\big)^{-1}A_{s}\psi_{s+1}\psi_{s+1}^\top A_{s},\;\;s=0,...,N-1.
\end{cases}
\eeq
\end{proposition}

\subsection{Robust BSL}
Even if the normality assumption of summary statistics is guaranteed, the standard BSL method might not be able to produce reliable inference about $\theta$
if the postulated model $\mathcal M$ is misspecified, in the sense that there is some considerable discrepancy between $\mathcal M$ and the underlying data-generating process that generated data $y_\text{obs}$.
This can happen if the summary statistics $s = s(\theta)$ cannot reproduce the behavior of the observed $s_\text{obs}$ for any $\theta$.
\cite{Frazier:JCGS2021} demonstrate that this mismatching issue might make BSL both statistically and computationally unstable.
To circumvent this problem, they propose a robust BSL (rBSL) method that adjusts the sample mean and variance for better matching.
To explain their method, we need some notation.
For a square matrix $A$, $\diag(A)$ denotes the diagonal vector of $A$.
For a vector $a$, $\diag(a)$ is the diagonal matrix formed by $a$,
and $a\circ b$ denotes the component-wise product of two vectors $a$ and $b$.
The rBSL method introduces a vector of auxiliary parameters $\Gamma=(\gamma_1,...,\gamma_d)^\top$, then defines the adjusted mean
\beq\label{eq:adjusted mean}
\wt\mu(\theta,\Gamma)=\wh\mu(\theta)+\diag(\wh\Sigma(\theta))^{1/2}\circ\Gamma.
\eeq
This adjustment requires working with $\wh\Sigma(\theta)$ which is the inverse of the precision $\wh P(\theta)$.
To avoid working with the sample covariance matrix, we propose a slight modification of \eqref{eq:adjusted mean} and work with 
\beq
\wt\mu(\theta,\Gamma)=\wh\mu(\theta)+\diag(\wh P(\theta))^{-1/2}\circ\Gamma.
\eeq
\cite{Frazier:JCGS2021} define their mean-adjusted robust BSL posterior as
\beq\label{eq: bsl-m posterior}
\overline\pi_\text{rBSL-M}(\theta,\Gamma)  \propto p(\theta)p(\Gamma)\overline{p}_\text{rBSL-M}(s_\text{obs}|\theta,\Gamma),
\eeq 
where $p(\Gamma)$ is the prior of $\Gamma$, and
\[\overline{p}_\text{rBSL-M}(s_\text{obs}|\theta,\Gamma)=\int \N_d\big(s_\text{obs};\wt\mu(\theta,\Gamma),\wh P(\theta)^{-1}\big)p(y_1,\dots, y_N|\theta)dy_1...dy_N.
\]
Note that, despite its notation, calculating the density $\N_d\big(s_\text{obs};\wt\mu(\theta,\Gamma),\wh P(\theta)^{-1}\big)$ only involves $\wh P(\theta)$.
Similarly, \cite{Frazier:JCGS2021} also define the variance-adjusted robust BSL but we will work with 
the adjusted-mean rBSL method for computational convenience.
As demonstrated in \cite{Frazier:JCGS2021}, the rBSL method can to some extent mitigate the model misspecification problem 
in likelihood-free Bayesian inference.

\subsection{Efficient VB for rBSL}
While using $\Gamma$ as the auxiliary parameters can address the problem with misspecified models, these auxiliary parameters might cause some difficulties in terms of model estimation.
\cite{Frazier:JCGS2021} use MCMC to sample from the joint posterior of 
$\Gamma$ and $\theta$, which can be computationally inefficient in complicated and high-dimensional applications. This paper uses Variational Bayes.
VB has proven a computationally attractive alternative to MCMC for statistical applications that involve demanding computations. See, e.g., \cite{Hoffman:JMLR2013,Blei:JASA2017,Loaiza-Maya:JoE2021} for the use of VB in various applications, and the reader is referred to \cite{Ong:STCO2018} and \cite{Tran:JCGS2017} for the previous use of VB in the likelihood-free context.

In modelling situations that involve auxiliary parameters, such as rBSL, a careful treatment of VB will lead to a more accurate and reliable VB approximation of the posterior of the main model parameters $\theta$ \citep{Loaiza-Maya:JoE2021,Dao:PM2022}. Recall the joint posterior of the main model parameter $\theta$ and auxiliary parameter $\Gamma$ in rBSL
\beq\label{eq: bsl posterior-nuissance}
\overline\pi_\text{rBSL}(\theta,\Gamma)  \propto \overline{p}_\text{rBSL}(\theta,\Gamma,s_\text{obs})=p(\theta)p(\Gamma)\overline{p}_\text{rBSL}(s_\text{obs}|\theta,\Gamma).
\eeq 
We propose a VB procedure that approximates the marginal 
\beqn
\overline\pi_\text{rBSL}(\theta)=\int \overline\pi_\text{rBSL}(\theta,\Gamma)d\Gamma\propto p(\theta)\overline{p}_\text{rBSL}(s_\text{obs}|\theta)=:\overline{p}_\text{rBSL}(\theta,s_\text{obs})
\eeqn
directly rather than approximating the joint $\overline\pi_\text{rBSL}(\theta,\Gamma)$; where $\overline{p}_\text{rBSL}(s_\text{obs}|\theta)=\int p(\Gamma)\overline{p}_\text{rBSL}(s_\text{obs}|\theta,\Gamma)d\Gamma$. 
It is shown in \cite{Dao:PM2022} and \cite{Tran:JCGS2017} that this {\it marginal} VB procedure leads to a 
more accurate approximation of the posterior distribution of $\theta$ in terms of Kullback-Leibler divergence.

More concretely, let $q_\lambda(\theta)$ be the variational approximation for $\overline\pi_\text{rBSL}(\theta)$, with $\lambda$ the variational parameters to be optimized.
The best $\lambda$ is found by maximizing the lower bound \citep{Blei:JASA2017}
\beqn
\LB(\lambda)=\E_{\theta\sim q_\lambda}\Big[\log\frac{\overline{p}_\text{rBSL}(\theta,s_\text{obs})}{q_\lambda(\theta)}\Big].
\eeqn
We note that the conditional posterior given $\theta$ of the auxiliary parameter $\Gamma$ is
\beq\label{eq: posterior of Gamma}
p(\Gamma|\theta,s_\text{obs}) = \frac{\overline{p}_\text{rBSL}(\theta,\Gamma,s_\text{obs})}{\overline{p}_\text{rBSL}(\theta,s_\text{obs})}.
\eeq
The gradient of the lower bound can be written as
\bea\label{eq: LB grad rBSL}
\nabla_\lambda\LB(\lambda)&=&\E_{\theta\sim q_\lambda}\Big[\nabla_\lambda\log q_\lambda(\theta)\circ\Big(\log\frac{\overline{p}_\text{rBSL}(\theta,s_\text{obs})}{q_\lambda(\theta)}-c\Big)\Big]\notag\\
&=&\E_{\theta\sim q_\lambda,\Gamma\sim p(\Gamma|s_\text{obs},\theta)}\Big[\nabla_\lambda\log q_\lambda(\theta)\circ\Big(\log\frac{\overline{p}_\text{rBSL}(\theta,s_\text{obs})}{q_\lambda(\theta)}-c\Big)\Big]\notag\\
&=&\E_{\theta\sim q_\lambda,\Gamma\sim p(\Gamma|s_\text{obs},\theta)}\Big[\nabla_\lambda\log q_\lambda(\theta)\circ\Big(\log\frac{\overline{p}_\text{rBSL}(\theta,\Gamma,s_\text{obs})}{q_\lambda(\theta)p(\Gamma|\theta,s_\text{obs})}-c\Big)\Big],
\eea
where the last equality is derived based on the equality in \eqref{eq: posterior of Gamma}, and $c$ is the vector of control variates \citep{Tran:JCGS2017}.
The derivation in \eqref{eq: LB grad rBSL} allows us to avoid 
working with the intractable term $\overline{p}_\text{rBSL}(\theta,s_\text{obs})$ by exploiting the role of the auxiliary parameter $\Gamma$.
\cite{Loaiza-Maya:JoE2021} and \cite{Dao:PM2022} also obtain expressions of the lower bound gradient similar to \eqref{eq: LB grad rBSL} when they approximate the joint posterior $p(\theta,\alpha|y)$ of a main model parameter $\theta$ and an auxiliary variable $\alpha$ by the VB approximation of the form $q_\lambda(\theta)p(\alpha|y,\theta)$.

Below we will work with the mean-adjusted rBSL and the prior $p(\Gamma)\sim \N(0,\sigma_0^2I_d)$. The posterior $p(\Gamma|\theta,s_\text{obs})$ becomes a Gaussian distribution $\N_d\big(\mu_\Gamma,\Sigma_\Gamma\big)$
with covariance matrix
\[\Sigma_\Gamma = \Big(\frac{1}{\sigma_0^2}I_d+\diag\big(\diag(\wh P(\theta))^{-1/2}\big)\wh P(\theta)\diag\big(\diag(\wh P(\theta))^{-1/2}\big)\Big)^{-1},\]
and mean
\[\mu_\Gamma = \Sigma_\Gamma\diag\big(\diag(\wh P(\theta))^{-1/2}\big)\wh P(\theta)(s_\text{obs} - \wh\mu(\theta)).\]
\cite{Frazier:JCGS2021} used a Laplace prior for $\Gamma$,
which encourages the latent $\gamma_j$ to concentrate on zero and only deviate from it if necessary.
However, this prior does not lead to a standard distribution for $p(\Gamma|\theta,s_\text{obs})$, which complicates the estimation 
of the lower bound gradient in \eqref{eq: LB grad rBSL}. 

The gradient of the lower bound becomes
\beq\label{eq:rBSL-M gradient}
\begin{aligned}
\nabla_\lambda\LB(\lambda)=&\E_{\theta\sim q_\lambda,y_{1:N}\sim p(\cdot|\theta),\Gamma\sim p(\Gamma|\theta,s_\text{obs})}\Big[\nabla_\lambda \log q_\lambda(\theta)\circ \Big(\log p(\theta)+\log\phi(\Gamma;0,\sigma_0^2I_d)+\\
&\log\phi(s_\text{obs};\wt\mu(\theta,\Gamma),\wh P(\theta)^{-1})-\log q_\lambda(\theta)-\log \phi(\Gamma;\mu_\Gamma,\Sigma_\Gamma)-c\Big)\Big],
\end{aligned}
\eeq
where $\phi(\cdot;\mu,P^{-1})$ denotes the multivariate normal density with mean $\mu$ and covariance matrix $\Sigma=P^{-1}$. 
We note that computing this density does not require the inverse $P^{-1}$, i.e.,
\[\log\phi(x;\mu,P^{-1})=-\frac{d}{2}\log(2\pi)+\frac12\log|P|-\frac12(x-\mu)^\top P(x-\mu).\]
We can obtain an unbiased estimator of $\nabla_\lambda\LB(\lambda)$ by first sampling $\theta \sim q_\lambda$, simulating data $y_{1:N}\sim p(\cdot|\theta)$ and then $\Gamma\sim p(\Gamma|\theta,s_\text{obs})$. This is summarized in Algorithm \ref{est gradient rbsl}. 

\begin{algorithm}[Estimated LB Gradient for rBSL]\label{est gradient rbsl}\phantom{c}
\begin{itemize}
\item Generate $S$ samples $\theta_i\sim q_\lambda(\cdot)$, $i=1,...,S$.
\item For each $\theta_i$, $i=1,...,S$, simulate $N$ datasets $y_{j}^{(i)}\sim p(\cdot|\theta_i)$, $j=1,...,N$, and generate $\Gamma_i\sim p(\Gamma|\theta_i,s_\text{obs})$.
 Let  $s_{j}^{(i)}$, $j=1,...,N$, be the corresponding summary statistics. Calculate the sample mean $\wh\mu(\theta_i)=\frac1N\sum_{j=1}^N s_j^{(i)}$
 and precision $\wh P(\theta_i)$ as in Lemma \ref{lemma}, and the adjusted mean:
\[\wt\mu(\theta_i,\Gamma_i)=\wh\mu(\theta_i)+\diag(\wh P(\theta_i))^{-1/2}\circ\Gamma_i.\]
\item The unbiased estimate of the lower bound gradient is
\[\wh{\nabla_\lambda\LB}(\lambda)=\frac{1}{S}\sum_{i=1}^S\nabla_\lambda \log q_\lambda(\theta_i)\circ \big(h_\lambda(\theta_i)-c\big),\]
where
\[ h_\lambda(\theta_i)=\log p(\theta_i)+\log p(\Gamma_i)+\log\phi(s_\text{obs};\wt\mu(\theta_i,\Gamma_i),\wh P(\theta_i)^{-1})-\log q_\lambda(\theta_i)-\log p(\Gamma_i|\theta_i,s_\text{obs}).\]
\end{itemize}
\end{algorithm}
From \cite{Tran:JCGS2017}, the optimal control variates $c=(c_1,...,c_D)$, with $D$ the length of $\lambda$, are 
\beq\label{eq:optimal c_i}
c_i=\cov\Big(\nabla_{\lambda_i}[\log q_\lambda(\t)]h_\lambda(\t),\nabla_{\lambda_i}[\log q_\lambda(\t)]\Big)\Big/\V\Big(\nabla_{\lambda_i}[\log q_\lambda(\t)]\Big),\ i=1,...,D,
\eeq
which can be estimated based on the samples $\theta_1,...,\theta_S$.

In this paper, we use Cholesky Gaussian VB where $q_\lambda(\theta)=\N(\mu,\Sigma)$ with $\Sigma^{-1}=CC^\top$ and $C$ a lower triangular matrix.
The variational parameter $\lambda$ is $\lambda=(\mu^\top,\vech(C)^\top)^\top$, and
\[\log q_\lambda(\theta)=-\frac{d}{2}\log(2\pi)+\log|C|-\frac{1}{2}(\theta-\mu)^\top CC^\top(\theta-\mu),\]
\[\nabla_\lambda\log q_\lambda(\theta)=
\begin{pmatrix}
CC^\top(\theta-\mu)\\
\vech\big(\diag(C^{-1})-(\theta-\mu)(\theta-\mu)^\top C)\big)
\end{pmatrix}.
\]
Algorithm \ref{algorithm 3} provides a detailed pseudo-code implementation of this CGVB approach that uses the control variate for variance reduction 
and moving average adaptive learning.

\begin{algorithm}[VB for rBSL]\label{algorithm 3} 

{\bf Input}: Initial $\l^{(0)}=(\mu^{(0)},C^{(0)})$, adaptive learning weights $\beta_1,\beta_2\in(0,1)$, fixed learning rate $\eps_0$, threshold $\tau$, rolling window size $t_W$ and maximum patience $P$. {\bf Model-specific requirement}: a mechanism to generate simulated data.
\begin{itemize}
  \item Initialization
  \begin{itemize}
	  \item Generate $\theta_s\sim q_{\lambda^{(0)}}(\theta)$, $s=1,...,S$ and compute the unbiased estimate of the LB gradient, $\wh{\nabla_\l\text{LB}}(\l^{(0)})$, as in Algorithm \ref{est gradient rbsl}.
	  \item Set $g_0:=\wh{\nabla_\l\text{LB}}(\l^{(0)})$, $v_0:=(g_0)^2$, $\bar g:=g_0$, $\bar v:=v_0$. 
	  \item Estimate the vector of control variates $c$ as in \eqref{eq:optimal c_i} using the samples $\{\theta_s,s=1,...,S\}$.
	  \item Set $t=0$, $\text{patience}=0$ and \texttt{stop=false}.
  \end{itemize}
  \item While \texttt{stop=false}:
  \begin{itemize}
	  \item Calculate $\mu^{(t)}$ and $C^{(t)}$ from $\lambda^{(t)}$. Generate $\theta_s\sim q_{\lambda^{(t)}}(\theta)$, $s=1,...,S$.
	  \item Compute the unbiased estimate of the LB gradient $g_t:=\wh{\nabla_\l\text{LB}}(\l^{(t)})$ as in Algorithm \ref{est gradient rbsl}.
	  \item Estimate the new control variate vector $c$ as in \eqref{eq:optimal c_i} using the samples $\{\theta_s,s=1,...,S\}$.
	  \item Compute $v_t=(g_t)^2$ and 
	  \[\bar g =\beta_1 \bar g+(1-\beta_1)g_t,\;\;\bar v =\beta_2 \bar v+(1-\beta_2)v_t.\]
	  \item Compute $\alpha_t=\min(\epsilon_0,\epsilon_0\frac{\tau}{t})$ and update
	  \[\l^{(t+1)}=\l^{(t)}+\a_t \bar g/\sqrt{\bar v}.\]
	  \item Compute the lower bound estimate
	  \[\wh{\text{LB}}(\l^{(t)}):=\frac{1}{S}\sum_{s=1}^Sh_{\lambda^{(t)}}(\theta_s).\]
	  \item If $t\geq t_W$: compute the moving averaged lower bound
	  \[\overline {\text{LB}}_{t-t_W+1}=\frac{1}{t_W}\sum_{k=1}^{t_W} \wh{\text{LB}}(\l^{(t-k+1)}),\]
	  and if $\overline {\text{LB}}_{t-t_W+1}\geq\max(\overline\LB)$ patience = 0; else $\text{patience}:=\text{patience}+1$.
\item If $\text{patience}\geq P$, \texttt{stop=true}.
\item Set $t:=t+1$.
  \end{itemize}
\end{itemize}
\end{algorithm}
In the numerical examples below,
we choose $\beta_1=\beta_2=0.9$, $t_W=P=50$,
$\tau=10,000$ and $\eps_0$ is in the range $0.001-0.01$.
The number of Monte Carlo samples $S$ is 1000,
and the number of simulated datasets $N$ in Algorithm \ref{est gradient rbsl} is typically 500.

\section{Wasserstein Gaussianization for Bayesian Synthetic Likelihood}\label{sec: WG for BSL}
The BSL method relies on the normality assumption on the summary statistics.
This section presents a method to fulfil this requirement, which constructs a flow of transformations
to transform the original summary statistics into a new vector whose distribution is approximately multivariate normal.
The method is based on the theory of Optimal Transport, which we describe next.

\subsection{Optimization on the Wasserstein space}
This section gives a brief introduction to the Wasserstein space of probability measures from Optimal Transport theory \citep{Ambrosio:OTbook,villani2009optimal}
and convex optimization on the Wasserstein space, which provides a foundation for our Wasserstein Gaussianization method.
Let $\mathcal P_2^\text{ac}(\SetR^d)$ be the space of probability measures on $\SetR^d$ with a finite second moment,
and absolutely continuous with respect to the Lesbegue measure $\mathcal L^d$ on $\SetR^d$.
With some abuse of notation, for a probability measure $\mu$, we use the notation $\mu(dx)$ to refer to it as a measure
and $\mu(x)$ for its density.
For any $\mu,\nu\in\mathcal P_2^\text{ac}(\SetR^d)$, let
\beq\label{eq:Wasserstein dis}
W_2(\mu,\nu)=\Big\{\inf_{T:T_{\#}\mu=\nu}\int_{\SetR^d}\|x-T(x)\|^2\mu(dx)\Big\}^{1/2}.
\eeq
Here, for a measurable map $T:\SetR^d\mapsto \SetR^d$, $T_{\#}\mu$ denotes the push-forward measure of $\mu$ through $T$, i.e. $T_{\#}\mu(A)=\mu(T^{-1}(A))$ for every measurable set $A\subset\SetR^d$.
Given that $\mu\ll \mathcal L^d$, it is well known that the {\it optimal transport map} $T$ that optimizes the integral in \eqref{eq:Wasserstein dis} exists and is unique $\mathcal L^d$-almost surely \citep{Ambrosio:OTbook,villani2009optimal}.
The quantity $W_2(\cdot,\cdot)$ is a metric on $\mathcal P_2^\text{ac}(\SetR^d)$; equipped with this metric, $\mathcal P_2^\text{ac}(\SetR^d)$ becomes a metric space, called the {\it Wasserstein space} and denoted by $\mathbb{W}_2(\SetR^d)$. 
This Wasserstein space has an attractive differential structure with a rich geometry 
that can be exploited to design efficient sampling and optimization techniques.    
Calculus concepts such as gradient, Hessian, geodesic convexity and smoothness of functionals on $\mathbb{W}_2(\SetR^d)$ are well defined.
We provide a review of these concepts in Appendix A.

Let $\pi\in\mathcal P_2^\text{ac}(\SetR^d)$ be a target probability measure that we want to approximate or sample from. 
Denote by $F(\mu)=\KL(\mu\|\pi)$ the KL divergence from a probability measure $\mu$ to $\pi$.
As 
\[\pi=\arg\min_{\mu\in \mathbb{W}_2(\SetR^d)}\big\{F(\mu)=\KL(\mu\|\pi)\big\},\] 
the problem of approximating $\pi$ can be cast as the problem of optimizing $F(\mu)$ over $\mathbb{W}_2(\SetR^d)$.
Suppose that $V(x)=-\log\pi(x)$ is $\alpha$-convex on $\SetR^d$, $\alpha\geq 0$; this assumption is satisfied in our Gaussianization problem below with $\alpha=1/2$, as the target $\pi$ is the standard multivariate Gaussian.
Then, it can be shown that $F$ is geodesically $\alpha$-convex on $\mathbb{W}_2(\SetR^d)$ \citep{Ambrosio:OTbook}[Chapter 9.4].
This guarantees that the minimizer of $F$ is unique and identical to the target $\pi$.

For a generic functional $G$ on $\mathbb{W}_2(\SetR^d)$, a typical method for optimizing $G$ is Wasserstein gradient descent.
Given some initial measure $\mu^{(0)}\in\mathcal P_2^\text{ac}(\SetR^d)$, for $k=0,1,...$,
\beq\label{eq:Wasserstein gradient}
\mu^{(k+1)} =T_{\#}\mu^{(k)},\;\;\;\text{ with }\;\;\;T(x)=x-\eps\nabla_\mu G(x),
\eeq
where $\nabla_\mu G:\SetR^d\mapsto \SetR^d$ is the Wasserstein gradient of $G$ at $\mu$ (see Appendix A for the definition), and $\epsilon>0$ is the step size. 
If $G$ is strongly geodesically convex and smooth, then the 
Wasserstein gradient method \eqref{eq:Wasserstein gradient} converges exponentially fast, i.e. $W_2^2(\mu^{(k)},\pi)=O(e^{-ck})$ for some constant $c>0$.
The Wasserstein gradient of the KL functional $F$ at $\mu$ is $\nabla_\mu F(x)=\nabla\big(\log{\mu(x)}-\log{\pi(x)}\big)$.


However, optimizing the KL functional $F$ over $\mathbb{W}_2(\SetR^d)$ is a challenging problem because, without further constraint on $\mu$, $F(\mu)$ is in general non-smooth.
More precisely, let us write $F$ as 
\[F(\mu)=\mathcal V(\mu)+\mathcal H(\mu),\]
where $\mathcal V(\mu)=\int V(x)d\mu(x)$, and $\mathcal H(\mu)=\int\log(\mu(x))d\mu(x)$ is the negative entropy.
The Hessian of $\mathcal H(\mu)$ is not bounded from above \citep{villani2009optimal}[Chapter 15],
therefore $\mathcal H(\mu)$ is not smooth. As a result, Wasserstein gradient descent for optimizing $F$ is not theoretically guaranteed to converge.

In this paper, we consider the proximal optimization scheme of \cite{Jordan:SIAM1998}. With some initial measure $\mu^{(0)}\in\mathcal P_2^\text{ac}(\SetR^d)$, at step $k\geq0$ and for a step size $\epsilon>0$, let
\beq\label{eq:JKO}
\mu^{(k+1)}:=\arg\min_{\mu\in\mathbb{W}_2(\SetR^d)}\Big\{F(\mu)+\frac{1}{2\eps}W_2^2(\mu,\mu^{(k)})\Big\}.
\eeq
This is often known as the JKO scheme in the literature. This scheme is an extension of the proximal algorithm \citep{parikh2014proximal} from the Euclidean space to the Wasserstein space.
\cite{Jordan:SIAM1998} show that, for each $k$, $\mu^{(k+1)}$ in \eqref{eq:JKO} exists and is unique.
The striking result obtained by \cite{Jordan:SIAM1998} is the connection between the JKO scheme, as $\epsilon\to0$, with the 
theory of gradient flows in the Wasserstein space.
They show that the discrete solution $\{\mu^{(k)}\}_k$ converges as $\eps\to0$ to a curve $\{\mu_t\}_{t\geq0}$ with the property that  
\beqn
\frac{{\d} F(\mu_t)}{{\d} t}=-\E_{\mu_t}\big(\|\nabla_{\mu_t} F\|^2\big)\leq 0.
\eeqn
That is, this curve optimizes the target functional $F(\mu)$, and is called the {\it Wasserstein gradient flow} of the KL functional $F(\mu)=\KL(\mu\|\pi)$ on $\mathbb{W}_2(\SetR^d)$.

For a fixed $\epsilon$, though, we show below that the JKO proximal scheme converges to the target $\pi$ with an exponential rate.
\begin{proposition}\label{pro:JKO convergence}
Suppose that $V(x)=-\log\pi(x)$ is $\alpha$-convex, $\alpha>0$.
The JKO proximal scheme converges exponentially fast in both functional value and Wassertein distance.
That is, for any initial measure $\mu^{(0)}$ such that $F(\mu^{(0)})<\infty$ and for any $k\geq0$,
\bea\label{eq:JKO proximal convergence 1}
F(\mu^{(k)})&\leq& (1+\epsilon\alpha)^{-k}F(\mu^{(0)}),\\
W_2^2(\mu^{(k)},\pi)&\leq& \frac{2}{\alpha}(1+\epsilon\alpha)^{-k}F(\mu^{(0)}).
\eea
\end{proposition}
In our Wasserstein Gaussianization algorithm below, $\alpha=1/2$.
The proof of Proposition \ref{pro:JKO convergence} 
can be found in Appendix B.

\subsection{Gaussianization by Wasserstein proximal optimization with normalizing flows}
This section describes a method for solving the optimization problem \eqref{eq:JKO}.
By Brenier's theorem \citep{brenier1991polar}, the solution $\mu^{(k+1)}$ must be of the form $\mu^{(k+1)}=(\nabla \varphi)_{\#}\mu^{(k)}$ where $\varphi(x)$ is a convex function.
We therefore can equivalently write \eqref{eq:JKO} as
\beq\label{eq:JKO new}
\mu^{(k+1)}=T_{\#}^{(k)}\mu^{(k)},\;\;  T^{(k)} = \arg\min_{T\in\mathcal{K}(\SetR^d)}\Big\{F(T) = \KL\big(T_{\#}\mu^{(k)}\|\pi\big)+\frac{1}{2\eps}\E_{\mu^{(k)}}\big(\|x-T(x)\|^2\big)\Big\},
\eeq
where $\mathcal{K}(\SetR^d)$ denotes the set of all transformations that are gradients of convex functions on $\SetR^d$.  

Following \cite{TsengTranNguyen:2024}, we parameterize the set $\mathcal{K}(\SetR^d)$ by a normalizing flow (NF).
NFs are transformations designed in such a way that their Jacobian is easy to compute \citep{papamakarios2021normalizing}, making them 
computationally efficient for solving \eqref{eq:JKO new}.
However, not all normalizing flows are gradients of convex functions.
The following lemma sets the conditions for a transformation to be the gradient of a convex function.
The proof is standard in calculus, hence omitted.
\begin{lemma}
A transformation $T:\mathbb{R}^d\to\mathbb{R}^d$ is the gradient of a convex function if and only if its Jacobian $\frac{\partial T(x)}{\partial x}$ is symmetric and non-negative definite.
\end{lemma}
With some appropriate conditions, many normalizing flows in the literature can be made into 
the gradients of convex functions. 
We refer to such NFs as {\it convex gradient NFs}.
We provide an example below; see \cite{TsengTranNguyen:2024} for more examples.

\paradot{Convex gradient Radial flow} Convex gradient radial flow has the following form
\begin{equation}\label{eq:Radial flow}
    y = T(x) = x+\frac{\gamma-\alpha}{\alpha+r(x)}(x-\mu),
\end{equation}
where $r(x)=\|x-\mu\|_2$, $\alpha>0$, $\gamma>0$ and $\mu\in\SetR^d$.
The Jacobian,
\begin{equation}
    \nabla T(x)=\frac{\partial T(x)}{\partial x}=\frac{\gamma+r(x)}{\alpha+r(x)}\mathbb{I}-\frac{\gamma-\alpha}{r(x)(\alpha+r(x))^2}(x-\mu)(x-\mu)^\top,
\end{equation}
is symmetric. Its determinant is 
\begin{equation}
    \det\left(\frac{\partial T(x)}{\partial x}\right) = \frac{\alpha\gamma+2\alpha r(x)+r(x)^2}{(\alpha+r(x))^2}\Big(\frac{\gamma+r(x)}{\alpha+r(x)}\Big)^{d-1}.
\end{equation}
For any vector $a\in\SetR^d$, $a\not=0$, it it easy to check that $a^\top\frac{\partial T(x)}{\partial x} a>0$. This shows that the transform $T(x)$ in \eqref{eq:Radial flow} is the gradient of a convex function. The parameters of this NF include $\lambda=(\alpha, \gamma, \mu)$. 

We now discuss how to solve \eqref{eq:JKO new} when $T$ is modelled by a NF.
Let $T=T_\lambda$ be a convex gradient NF with parameters $\lambda$.
Write $\nu=(T_\lambda)_{\#}\mu^{(k)}$ for the measure obtained from $\mu^{(k)}$ through transformation $T_\lambda$. By the change of variables formula,
\[\nu(T_\lambda(x))=\bigg|\frac{\partial T_\lambda(x)}{\partial (x)}\bigg|^{-1}\mu^{(k)}(x).\]
Then,
\bean
\KL\big((T_\lambda)_{\#}\mu^{(k)}\|\pi\big)&=&\E_{\nu}\left(\log\frac{\nu(x)}{\pi(x)}\right)=\int \log\frac{\nu(x)}{\pi(x)} {\d}(T_\lambda)_{\#}\mu^{(k)} \\
&=&\int \log\frac{\nu(T_\lambda(x))}{\pi(T_\lambda(x))} {\d}\mu^{(k)}\\
&=&\int\Big(-\log\big|\frac{\partial T_\lambda(x)}{\partial (x)}\big|+\log\big(\mu^{(k)}(x)\big)-\log\pi\big(T_\lambda(x)\big)\Big){\d}\mu^{(k)}
\eean
The objective function $F(T)$ in \eqref{eq:JKO new} becomes
\bean
F(T_\lambda)&=&\E_{\mu^{(k)}}\Big(-\log\big|\frac{\partial T_\lambda(x)}{\partial (x)}\big|+\log\big(\mu^{(k)}(x)\big)-\log\pi\big(T_\lambda(x)\big)\Big)+\frac{1}{2\eps}\E_{\mu^{(k)}}\big(\|x-T_\lambda(x)\|^2\big)\\
&=&\E_{\mu^{(k)}}\Big(-\log\big|\frac{\partial T_\lambda(x)}{\partial (x)}\big|-\log\pi\big(T_\lambda(x)\big)+\frac{1}{2\eps}\|x-T_\lambda(x)\|^2\Big)+C,
\eean
where the constant $C=\E_{\mu^{(k)}}\big(\log\big(\mu^{(k)}(x)\big)$
is independent of $\lambda$. Minimizing $F(T_\lambda)$ in $\lambda$ is therefore equivalent to minimizing
\beq\label{eq:objective function}
F(\lambda):=\E_{\mu^{(k)}}\Big(-\log\bigg|\frac{\partial T_\lambda(x)}{\partial (x)}\bigg|-\log\pi\big(T_\lambda(x)\big)+\frac{1}{2\eps}\|x-T_\lambda(x)\|^2\Big).
\eeq
As we already have samples from measure $\mu^{(k)}$, it is a standard task to optimize $F(\lambda)$ using stochastic gradient descent (SGD). Let $\lambda^{(k)}$ be a minimizer of $F(\lambda)$, the next distribution is $\mu^{(k+1)}=(T_{\lambda^{(k)}})_{\#}(\mu^{(k)})$.

For Gaussianization, the target $\pi$ will be the standard Gaussian $\mathcal N(0,I)$, i.e. $\pi(x)\propto\exp(-\frac12x^\top x)$.
The resulting Gaussianization transformation $T$ is the composition of the $T_{\lambda^{(k)}}$.
More specifically, let $\{x_i^{(0)}\}_{i=1}^M$ be samples generated from an initial measure $\mu^{(0)}$.
At step $k\geq0$, let $\lambda^{(k)}$ be the minimizer of $F(\lambda)$ in \eqref{eq:objective function}. 
Then, we move the particles as
\[x^{(k+1)}_i=T_{\lambda^{(k)}}(x^{(k)}_i),\;\;\;i=1,...,M\]
and set $T\gets T_{\lambda^{(k)}}\circ T$.

\paradot{Stopping rule}
We now discuss the stopping rule.
Denote $T^{(k)}=T_{\lambda^{(k)}}\circ T_{\lambda^{(k-1)}}\circ \cdots \circ T_{\lambda^{(0)}}$ be the transformation obtained after step $k$.
The $M$ particles are traversed via 
\[x^{(k+1)}_i=T_{\lambda^{(k)}}(x^{(k)}_i)=T^{(k)}(x_i^{(0)}),\;\;\;i=1,...,M.
\]
For a generic particle $x^{(0)}$, it will be helpful to write
\bean
x^{(1)}&=&T^{(0)}(x^{(0)})=T_{\lambda^{(0)}}(x^{(0)})\\
x^{(2)}&=&T^{(1)}(x^{(0)})=T_{\lambda^{(1)}}(x^{(1)})\\
...\\
x^{(k+1)}&=&T^{(k)}(x^{(0)})=T_{\lambda^{(k)}}(x^{(k)}).
\eean
The particles $\{x_i^{(k+1)}\}_{i=1}^M$ are distributed with respect to the probability measure with density
\[\mu^{(k+1)}(x^{(k+1)})=\big|\frac{\partial T^{(k)}(x^{(0)})}{\partial x^{(0)}}\big|^{-1}\mu^{(0)}(x^{(0)}),\]
where the determinant term can be written as
\bean
\big|\frac{\partial T^{(k)}(x^{(0)})}{\partial x^{(0)}}\big|&=&
\big|\frac{\partial x^{(k+1)})}{\partial x^{(k)}}\big|\times
\big|\frac{\partial x^{(k)})}{\partial x^{(k-1)}}\big|\times\cdots
\times \big|\frac{\partial x^{(1)})}{\partial x^{(0)}}\big|\\
&=&\big|\nabla T_{\lambda^{(k)}}(x^{(k)})\big|\times
\big|\nabla T_{\lambda^{(k-1)}}(x^{(k-1)})\big|\times\cdots\times \big|\nabla T_{\lambda^{(0)}}(x^{(0)})\big|.
\eean
To assess the stopping rule, we use the KL divergence from $\mu^{(k+1)}$ to the target $\pi$
\bean
\KL\big(\mu^{(k+1)}\|\pi\big)&=&\E_{\mu^{(k+1)}}\Big(\log\frac{\mu^{(k+1)}(x^{(k+1)})}{\pi(x^{(k+1)})}\Big)\\
&=&\E_{\mu^{(0)}}\Big(\log\frac{\mu^{(k+1)}\big(T^{(k)}(x^{(0)})\big)}{\pi\big(T^{(k)}(x^{(0)})\big)}\Big)\\
&=&\E_{\mu^{(0)}}\Big(-\sum_{j=0}^k\log\big|\nabla T_{\lambda^{(j)}}(x^{(j)})\big|+\log\mu^{(0)}(x^{(0)})-\log\pi\big(T^{(k)}(x^{(0)})\big)\Big).
\eean
The term $\E_{\mu^{(0)}}\Big(\log\mu^{(0)}(x^{(0)})\Big)$ is unknown, but it is a constant across the iteration, thus can be ignored.
The term
\bea\label{eq: Lower bound}
\LB_k &:=& \E_{\mu^{(0)}}\Big(\sum_{j=0}^k\log\big|\nabla T_{\lambda^{(j)}}(x^{(j)})\big|+\log\pi\big(T^{(k)}(x^{(0)})\big)\Big)\notag\\
&=&-\frac{d}{2}\log(2\pi)+\E_{\mu^{(0)}}\Big(\sum_{j=0}^k\log\big|\nabla T_{\lambda^{(j)}}(x^{(j)})\big|-\frac12\|T^{(k)}(x^{(0)})\|^2\Big),
\eea
should be increasing over the iteration $k$, and we refer to it as the lower bound. 
This term can be estimated using the particles $\{x_i^{(0)}\}_{i=1}^M$, and used for the stopping rule. 

We summarize our proposed Wasserstein Gaussianization algorithm below.
\begin{algorithm}[Wasserstein Gaussianization with NF]\label{alg:Gaussianization WNF} 
Let $\{x_i^{(0)}\}_{i=1}^M$ be original samples generated from a measure $\mu^{(0)}$.
Initialize the transformation $T=\text{Id}$ and $\Delta:=-\frac{d}{2}\log(2\pi)$. For $k=0,1,...$, iterating until convergence: 
\begin{itemize}
\item Optimize $F(\lambda)$ in \eqref{eq:objective function} by SGD. Let $\lambda^{(k)}$ be its minimizer.
\item Compute 
\[\Delta\gets \Delta+\frac1M\sum_{i=1}^M\log\big|\nabla T_{\lambda^{(k)}}(x^{(k)}_i)\big|.\]
\item Compute the lower bound
\[\LB_k=\Delta-\frac1{2M}\sum_{i=1}^M\|x^{(k)}_i\|^2.\]
\item If $\LB_k$ increases: set $T\gets T_{\lambda^{(k)}}\circ T$ and move the particles
\[x^{(k+1)}_i=T_{\lambda^{(k)}}(x^{(k)}_i),\;\;\;i=1,...,M;
\]
otherwise, stop.
\end{itemize}
\end{algorithm}

\begin{remark}
In our application of WG in the BSL method below, the initial samples $\{x_i^{(0)}\}_{i=1}^M$ are the original summary statistics generated from the model at some given value of $\theta$.
With a minor modification, Algorithm \ref{alg:Gaussianization WNF} can be used to traverse $\{x_i^{(0)}\}_{i=1}^M$ to become samples from any target probability distribution $\pi$, such as a posterior distribution.
\end{remark}

\subsection{A toy example}\label{sec: toy example} 
Consider the data generating process
\beq\label{eq:DGP toy example}
y_i = \theta+\varepsilon_i,\;\;\;\;\theta\in\SetR;
\eeq
the $\varepsilon_i$ are i.i.d. random errors with mean 0 and variance $\sigma^2=4$. To create errors with a heavily skewed distribution, we set $\varepsilon_i=\sigma (v_i-\alpha/\beta)/\sqrt{\alpha/\beta^2}$, where $v_i$ follows a Gamma distribution with shape parameter $\alpha=1$ and rate $\beta=0.01$. For the observed dataset, we generate a set of $n=30$ observations $y_\text{obs}=\{y_i,\ i=1,...,n\}$ from model \eqref{eq:DGP toy example} with $\theta=0$. Given this dataset, we then proceed with Bayesian inference for the parameter $\theta$. Reasonable summary statistics are the sample mean and variance $s_\text{obs}=(\bar y,s^2)$.

\begin{figure}[ht]
\centering
\includegraphics[width=0.5\textwidth]{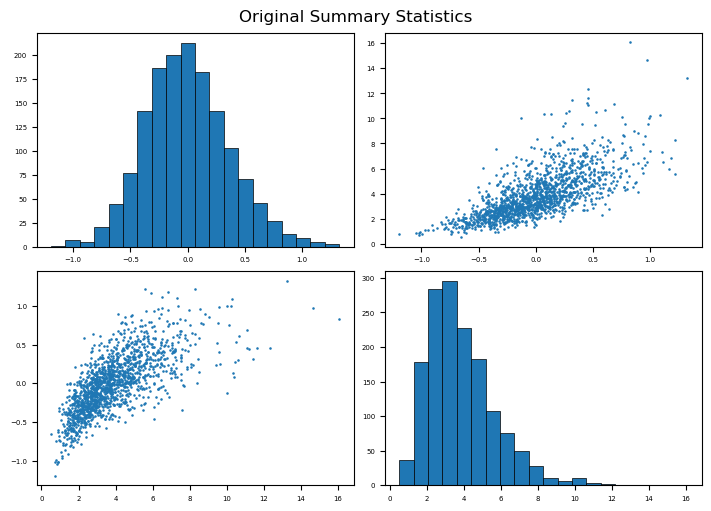}
\caption{Toy example: The scatter plots of the original summary statistics.}
\label{fig:toy example 1-nf}
\end{figure}

\begin{figure}[H]
     \centering
     \begin{subfigure}[b]{0.45\textwidth}
         \centering
         \captionsetup{labelformat=empty}
         \includegraphics[width=\textwidth]{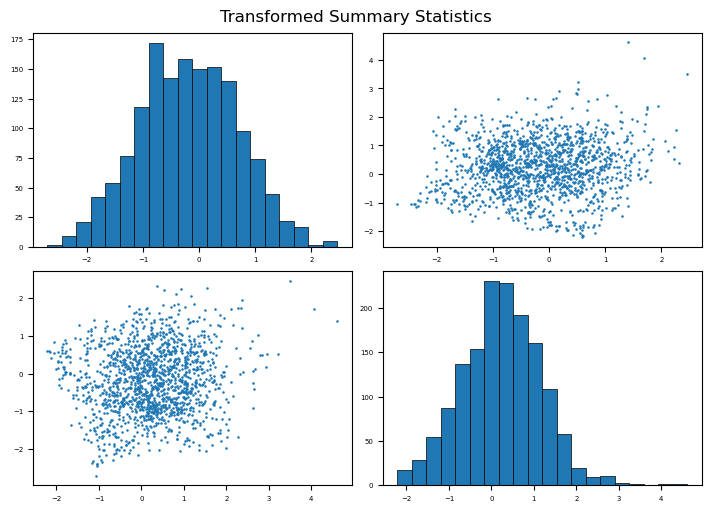}
     \end{subfigure}
     \begin{subfigure}[b]{0.45\textwidth}
         \centering
         \captionsetup{labelformat=empty}
         \includegraphics[width=\textwidth]{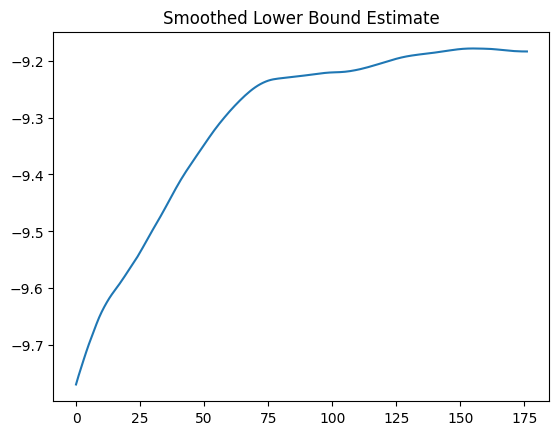}
     \end{subfigure}
     \caption{Toy example: The left panel shows the scatter plots of the transformed summary statistics. The right plot shows the LB.}\label{fig:toy_example_without tau}
\end{figure}

To examine if these summary statistics follow a normal distribution, we generate $M=10,000$ datasets, each of size $n=30$ with $\theta = 0$, and calculate the set of corresponding $M$ summary statistics. 
The scatter plots in \autoref{fig:toy example 1-nf}
indicate that the summary statistics do not follow a normal distribution.
Dividing these datasets into a training and validation set, we now run Algorithm \autoref{alg:Gaussianization WNF} for training a WG transformation.
The left panel of \autoref{fig:toy_example_without tau} shows the scatter plots of the transformed summary statistics on the validation set, which visually indicate that they follow a normal distribution; the right panel plots the LB estimate \eqref{eq: Lower bound} values over iterations. To formally assess the efficacy of these transformations in enhancing the multivariate normality of the original summary statistics, we utilize the Henze-Zirkler (HZ) multivariate normality test \citep{henze1990class}. 
The $p$-value for the HZ test with the original summary statistics is 0.004, 
while it is 0.0561 for the transformed summary statistics.
If we choose the typical significance level of 0.05, we then reject the normality of the original summary statistics while cannot do so for the 
transformed summary statistics.

\subsection{Robust BSL with Wasserstein Gaussianization}
As robust BSL still requires the normality assumption, it is natural to combine the WG method described in the previous section with robust BSL.
We refer to this method as rBSL-WG.
Let $T$ be the WG transformation trained at some central value $\theta_0$.
The rBSL-WG works with the following posterior
\beq\label{eq: rBSL-WG posterior}
\overline\pi_\text{rBSL-WG}(\theta,\Gamma)  \propto p(\theta)p(\Gamma)\overline{p}_\text{rBSL-WG}(s_\text{obs}|\theta,\Gamma),
\eeq 
where
\[\overline{p}_\text{rBSL-WG}(s_\text{obs}|\theta,\Gamma)=\int \N_d\big(s_\text{obs};\wt\mu_\text{WG}(\theta,\Gamma),\wh P_\text{WG}(\theta)^{-1}\big)p(y_1,\dots, y_N|\theta)dy_1...dy_N
\]
with the adjusted mean 
\[\wt\mu_\text{WG}(\theta,\Gamma)=\wh\mu_\text{WG}(\theta)+\diag(\wh P_\text{WG}(\theta))^{-1/2}\circ\Gamma.\]
Here, 
\beq\label{eq: mean and variance}
\wh\mu_\text{WG}(\theta)=\frac1N\sum_{j=1}^N T(s_j),\;\;\;\wh P_\text{WG}(\theta)=N\Big(\Psi_0+\sum_{j=1}^N(T(s_j)-\hat\mu_\text{WG}(\theta))(T(s_j)-\hat\mu_\text{WG}(\theta))^\top\Big)^{-1}.
\eeq
Note that the precision $\wh P_\text{WG}(\theta)$ is computed using the recursive procedure as in Lemma \ref{lemma}. That is, let $\psi_j=T(s_j)-\hat\mu_\text{WG}(\theta)$, then 
$\wh P_\text{WG}(\theta)=NA_N$ where
\beqn
\begin{cases}
A_0 = \Psi_0^{-1},\\
A_{s+1} = A_{s}-\big(1+\psi_{s+1}^\top A_{s}\psi_{s+1}\big)^{-1}A_{s}\psi_{s+1}\psi_{s+1}^\top A_{s},\;\;s=0,...,N-1.
\end{cases}
\eeqn
Our VB algorithm for rBSL-WG is the same as Algorithm \ref{algorithm 3}, except that the sample mean and sample precision matrix are replaced by $\wh\mu_\text{WG}(\theta_i)$ and
$\wh P_\text{WG}(\theta_i)$, respectively.

\section{Numerical examples and applications}\label{sec: examples}
This section implements and compares the performance of the four BSL methods using VB for inference: standard BSL (VB-BSL), robust BSL (VB-rBSL), 
standard BSL with Wasserstein Gaussianization (VB-BSL-WG) and robust BSL with Wasserstein Gaussianization (VB-rBSL-WG). The implementation was in Python and run on a Dell computer equipped with an Intel Core i7-1265U 1.80 GHz processor and 10 CPU cores. The code 
is executed using JAX, a powerful numerical computing library known for its scalability.
The computer code for the examples is available at \url{https://github.com/VBayesLab/VBSL-WG}.

\subsection{Toy example}
We consider again the toy example outlined in Section \ref{sec: toy example}. For each iteration in the VB Algorithm \ref{algorithm 3}, we generate $S=400$ samples of $\theta$, from each of which we simulate $N=200$ datasets each of size $n=200$. 
The algorithmic parameters for Wasserstein Gaussianization transformation $T$ are as in Section \ref{sec: toy example}.

\begin{figure}[H]
     \centering
     \begin{subfigure}[b]{0.45\textwidth}
         \centering
         \captionsetup{labelformat=empty}
         \includegraphics[width=\textwidth]{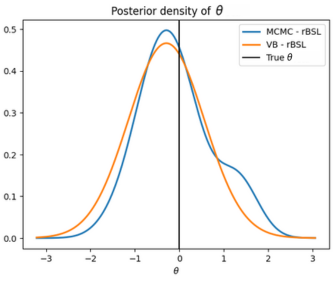}
     \end{subfigure}
     \begin{subfigure}[b]{0.45\textwidth}
         \centering
         \captionsetup{labelformat=empty}
         \includegraphics[width=\textwidth]{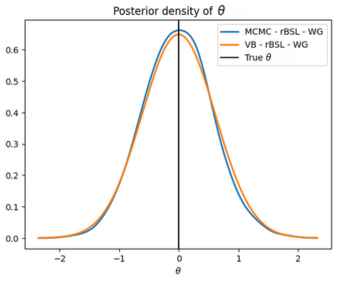}
     \end{subfigure}
     \caption{Toy example: VB v.s. MCMC posterior estimates of $\theta$.}\label{fig:toy_example_MCMCvsVB}
\end{figure}

Figure \ref{fig:toy_example_MCMCvsVB} plots the posterior estimates of $\theta$, which show that both VB and MCMC estimates are around the true parameter.
This is interesting as the rBSL-WG method is able to recover the true model parameter in this example, although there is no model misspecification. This suggests that the rBSL helps robustify and accommodate the normality assumption to some extent; see \cite{Frazier:JCGS2021} for more detailed discussion.

\begin{figure}[ht]
     \centering
     \begin{tabular}{@{}c@{}}
         \centering
         \includegraphics[width=0.5\textwidth]{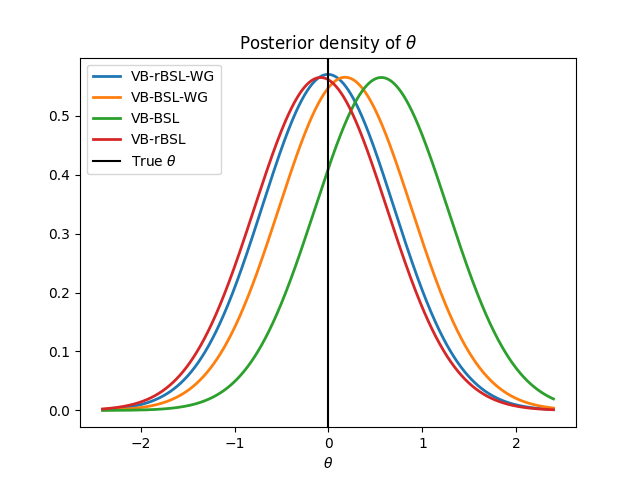}
     \end{tabular}
        \caption{Toy example: The plot shows variational posterior across the four BSL  methodologies}
        \label{fig:posterior-mean-theta-toy-nf}
\end{figure}

We evaluate the efficacy of incorporating WG transformations into different variations of VB-BSL via \autoref{fig:posterior-mean-theta-toy-nf} and \autoref{tab:l2norm toy example-nf}. \autoref{fig:posterior-mean-theta-toy-nf} shows the plot of the VB posterior estimates. The mean squared error (MSE) between the posterior mean estimate of $\theta$ and the true $\theta$ across 10 distinct runs is presented in \autoref{tab:l2norm toy example-nf}. This table contrasts various VB-BSL versions with and without WG transformations. \autoref{tab:l2norm toy example-nf} shows that adding the Wasserstein Gaussianization into the BSL method is useful and helps improve the estimation accuracy. Particularly, both rBSL and rBSL-WG exhibit significant improvements over the standard BSL. Moreover, VB-rBSL-WG outperforms VB-rBSL marginally. This might be because of the oversimplicity in this one-dimensional example; we observe larger improvements in the more sophisticated examples in the next sections.

\begin{table}[ht]
\centering
\begin{tabular}{|P{5em}|P{2.75cm}|P{2.75cm}|P{2.75cm}|P{2.75cm}|}
\hline
                                       & VB-BSL & VB-rBSL & VB-BSL-WG & VB-rBSL-WG \\ \hline
$||\theta_{true} - \hat{\theta}||_2$ & 0.0509 (0.0236) & 0.0835 (0.0148)  & 0.1780 (0.0259) & 0.0053 (0.0049)    \\ \hline
\end{tabular}
\caption{Toy example: The averaged MSE, over the replicates, of the parameter estimates across the four variations of BSL methods. The numbers in brackets are standard deviations.}
\label{tab:l2norm toy example-nf}
\end{table}

We further evaluate the robustness of the WG transformation method by assessing its performance across various methods of summarizing data. We explore three different summary statistics approaches. The first approach retains the use of $s_\text{obs}=(\bar y,s^2)$ as above. The second approach summarizes data using quartiles, $s_\text{obs}=(P_{25}, P_{50}, P_{75})$, where $P_j$ represents the $j$-th percentile of the data, and the third approach uses $s_\text{obs}=(P_{20}, P_{40}, P_{60}, P_{80})$. \autoref{tab:ss} shows the MSE values averaged over 10 runs for each summarization method, labeled as SS$_1$, SS$_2$, and SS$_3$. 
Compared to the standard VB-BSL, we observe negligible variations in the efficacy of VB-rBSL-WG across different summarization strategies in capturing the true model parameters. This indicates that incorporating the WG transformation enhances the robustness of BSL with respect to summary statistics.

\begin{remark}
Choosing appropriate summary statistics in likelihood-free inference is a difficult topic, and there is an extensive literature. Common approaches include projecting from a large candidate pool of summary statistics to a lower-dimensional space \citep{fearnhead2012constructing} or using auxiliary models \citep{drovandi2015bayesian}. For a survey, see \cite{prangle2018summary}. Additionally, the selection of summary statistics can sometimes be informed by domain experts for specific applications. In principle, our method is applicable regardless of the choice of the summary statistics, provided that the choice meets the assumptions outlined in \cite{sisson2018handbook}.  
\end{remark}

\begin{table}[]
\centering
\begin{tabular}{|P{5em}|P{2.15cm}P{2.15cm}|P{2.15cm}P{2.15cm}|P{2.15cm}P{2.15cm}|}
\hline
\multirow{2}{*}{}                    & \multicolumn{2}{c|}{SS$_1$}                                                                            & \multicolumn{2}{c|}{SS$_2$}                                                                                        & \multicolumn{2}{c|}{SS$_3$}                                                                             \\ \cline{2-7} 
                                     & \multicolumn{1}{c|}{VB-BSL}                                                    & \begin{tabular}[c]{@{}c@{}}VB-rBSL-\\ WG\end{tabular} & \multicolumn{1}{c|}{VB-BSL}                                                     & \begin{tabular}[c]{@{}c@{}}VB-rBSL-\\ WG\end{tabular} & \multicolumn{1}{c|}{VB-BSL}                                                     & \begin{tabular}[c]{@{}c@{}}VB-rBSL-\\ WG\end{tabular} \\ \hline
$||\theta_\text{true} - \hat{\theta}||_2$ & \multicolumn{1}{c|}{\begin{tabular}[c]{@{}c@{}}0.0509\\ (0.0236)\end{tabular}} & \begin{tabular}[c]{@{}c@{}}0.0059 \\ (0.008)\end{tabular}   & \multicolumn{1}{c|}{\begin{tabular}[c]{@{}c@{}}0.0328 \\ (0.0198)\end{tabular}} & \begin{tabular}[c]{@{}c@{}}0.0053\\ (0.0014)\end{tabular} & \multicolumn{1}{c|}{\begin{tabular}[c]{@{}c@{}}0.0588 \\ (0.0325)\end{tabular}} & \begin{tabular}[c]{@{}c@{}}0.0057\\ (0.0046)\end{tabular}   \\ \hline
\end{tabular}
\caption{Toy example: The averaged MSE, over the replicates, of the parameter estimates across the three summarization approaches. The numbers in brackets are standard deviations.}
\label{tab:ss}
\end{table}

\subsection{$\alpha$-stable model example}
This section applies the BSL methods to Bayesian inference in the $\alpha$-stable model \citep{Nolan.book:2007}, which is a family of heavy-tailed distributions that have seen applications in situations with rare but extreme events. This family of distributions poses challenges for inference because there exists no closed form expression for the density function. 
The $\alpha$-stable distribution is often defined using its log characteristic function \citep{samorodnitsky2017stable}
\[ \log\phi(t) = \begin{cases}
    -\gamma^\alpha |t|^\alpha {1 - i\beta\text{sign}(t)\tan\frac{\pi \alpha}{2}} + i\delta t, & \alpha \neq 1 \\
    -\gamma |t| {1 + i\beta\text{sign}(t)  \frac{2}{\pi}\log|t|} + i\delta t, & \alpha = 1,
\end{cases}
\]
where $\theta=(\alpha, \beta, \gamma, \delta)$ are the model parameters and $\text{sign}(t)$ is equal to 1 if $t\geq0$ and $-$1 if $t<0$. As for the observed data, we simulated a dataset of size $n=200$ with the true parameters $(\alpha, \beta, \gamma, \delta) = (1.8, 0.5, 1, 0)$. 

We now test the ability of the BSL methods to recover the true parameter together with the estimation uncertainty.
We follow \cite{Peters:2012} and enforce the constraints $\alpha \in (1.1,2)$, $\beta \in (-1,1)$, $\gamma > 0$, 
and work with the following reparametrization of $\theta$ 
\[\tilde{\alpha} = \log \frac{\alpha-1.1}{2-\alpha}, \quad \tilde{\beta} = \log \frac{\beta+1}{1-\beta}, \quad \tilde{\gamma} = \log \gamma, \quad \tilde{\delta} = \delta.\]
We use a normal prior $N(0,100I_4)$ for $\tilde{\theta}$ as in \cite{Tran:JCGS2017}.
We then approximate the posterior of $\tilde{\theta}=(\tilde{\alpha},\tilde{\beta},\tilde{\gamma},\tilde{\delta})$, but the results below will be reported in terms of the original parameter $\theta$.

\begin{figure}[ht]
     \centering
     \begin{tabular}{@{}c@{}}
         \centering
         \includegraphics[width=0.3\textwidth]{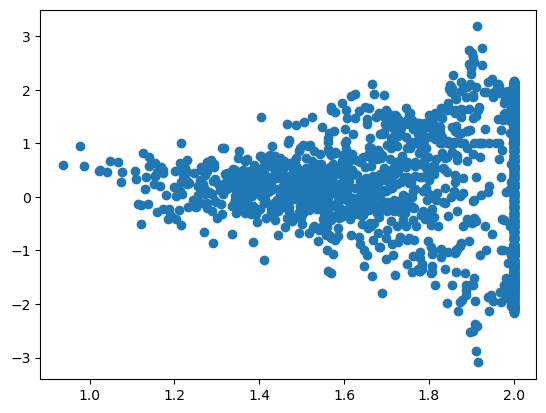}
     \end{tabular}
     \begin{tabular}{@{}c@{}}
         \centering
         \includegraphics[width=0.3\textwidth]{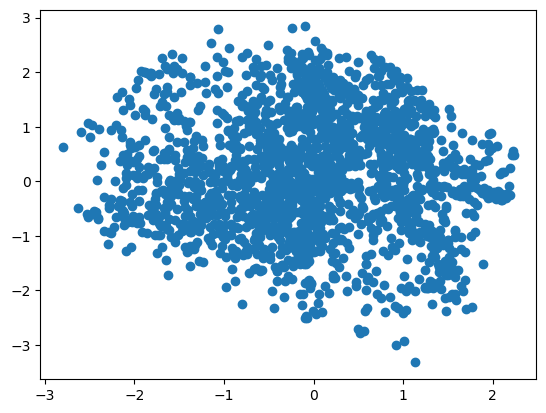}
     \end{tabular}
     \begin{tabular}{@{}c@{}}
         \centering
         \includegraphics[width=0.3\textwidth]{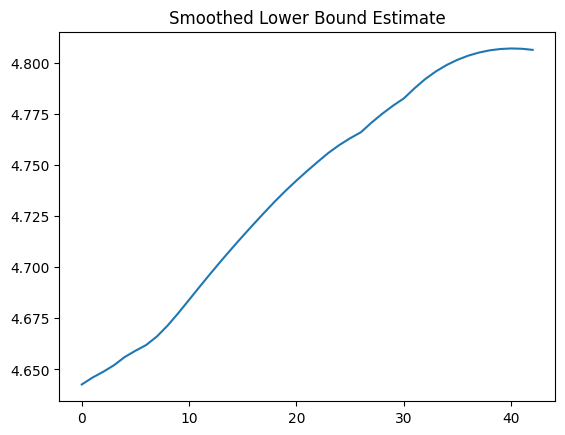}
     \end{tabular}
        \caption{$\alpha$-stable model: The scatter plots of the original and NF transformed summary statistics. The right plot shows the LB estimates. To save space, only one of the six possible pairs derived from the summary statistics is displayed.}
        \label{fig:alpha-stable-scatterplot-nf}
\end{figure}

For the Wasserstein Gaussianization, we first run the standard VB-BSL algorithm to get an estimate, $\theta_0$, of the posterior mean of $\theta$. We then generate 30,000 datasets, each comprising 30 observations, from the $\alpha$-stable distribution and compute the summary statistics.
For the four summary statistics used in the $\alpha$-stable distribution, we refer the interested readers to \cite{Peters:2012}.
We divide the 30,000 sets of summary statistics into a training and a validation set, and use the validation set to determine the adherence of the summary statistics to a multivariate normal distribution. Following the removal of outliers, we compute the $p$-value from the HZ multivariate normality test.
The $p$-value for the original summary statistics is $0.0074$, whereas for the transformed summary statistics, it is 0.1688. This disparity indicates that the original summary statistics do not satisfy the normality test, while the transformed ones do.
The left panel in \autoref{fig:alpha-stable-scatterplot-nf} displays a scatter plot of the original summary statistics, indicating departure from a normal distribution. The middle panel presents the scatter plot of the transformed summary statistics on the validation dataset, exhibiting a closer alignment with a normal distribution. Finally, the right panel illustrates the LB estimates. 
The computational time required for the WG transformation was approximately 59 minutes, which can be attributed to the complexity resulting from the use of 12 layers in NF, as well as the additional computational overhead incurred by the random permutation process used to select batches from the 10,000 datasets.

We proceed by executing the four BSL methods with VB as the estimation technique: BSL, rBSL, BSL-WG, rBSL-WG. Our objective is to assess the efficacy of these BSL methods in terms of their capability to recover the true model parameters. To have a better measure of the performance of the methods, we repeat this example ten times
and compute the average MSE and Mahalanobis distance between the posterior mean estimates and the true parameters. The Mahalanobis distance is calculated as 
\[\text{MD}(||\theta_{true} - \hat{\theta}||) = \sqrt{(\theta_\text{true} - \hat{\theta}){\hat\Sigma}^{-1}(\theta_{true} - \hat{\theta})^\top}
\]
where $\hat\theta$ and $\hat\Sigma$ are the mean and covariance matrix obtained from the optimal variational Gaussian distribution.
The MSE, Mahalanobis distance values and CPU times are summarized in Table \ref{tab:l2alpha-nf}.

\begin{table}[ht]
\centering
\resizebox{\textwidth}{!}{\begin{tabular}{|P{7em}|P{2.75cm}|P{2.75cm}|P{2.75cm}|P{2.75cm}|}
\hline
                                       & VB-BSL & VB-rBSL & VB-BSL-WG & VB-rBSL-WG \\ \hline
$||\theta_\text{true} - \hat{\theta}||_2$ & 0.3182 (0.009) & 0.0167 (0.002)  & 0.0461 (0.0157) & 0.0039 (0.0021)     \\ \hline 
MD$(||\theta_\text{true} - \hat{\theta}||)$ & 3.1967 (0.0359) & 0.1680 (0.0023)  & 0.4284 (0.1108)    & 0.0318 (0.0235)     \\ \hline
CPU time (min) & 9.1 & 8.9  & 21.0    & 36.2 \\
\hline
\end{tabular}}
\caption{$\alpha$-stable model: MSE of the posterior mean estimates compared to $\theta_{\text{true}}$, over 10 different runs. The numbers in brackets are standard deviations.}
\label{tab:l2alpha-nf}
\end{table}

\begin{figure}[h]
     \centering
     \begin{tabular}{@{}c@{}}
         \centering
         \includegraphics[width=0.35\textwidth]{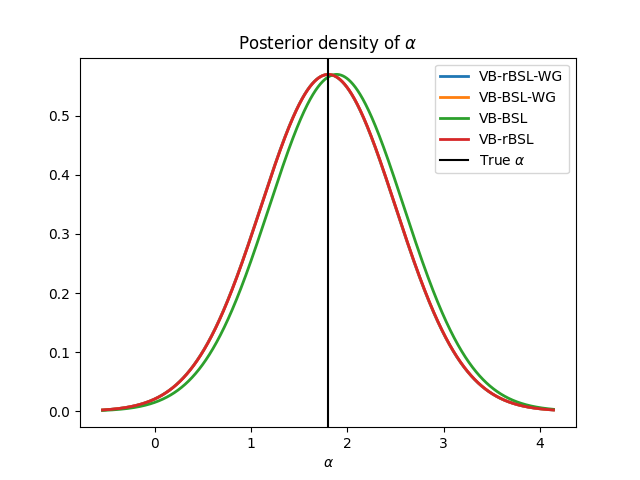}
     \end{tabular}
     \begin{tabular}{@{}c@{}}
         \centering
         \includegraphics[width=0.35\textwidth]{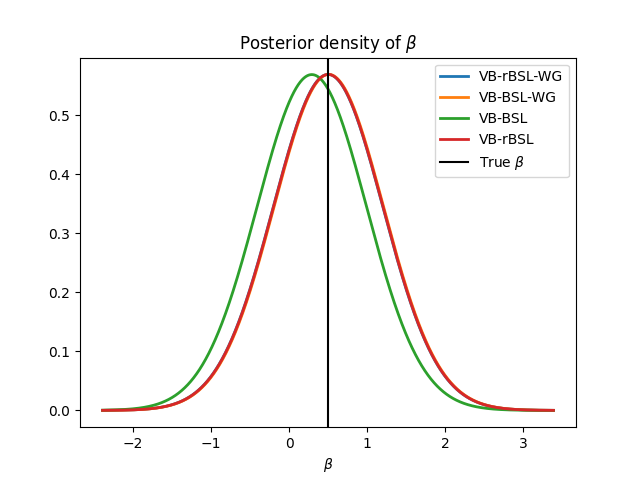}
     \end{tabular}
     \begin{tabular}{@{}c@{}}
         \centering
         \includegraphics[width=0.35\textwidth]{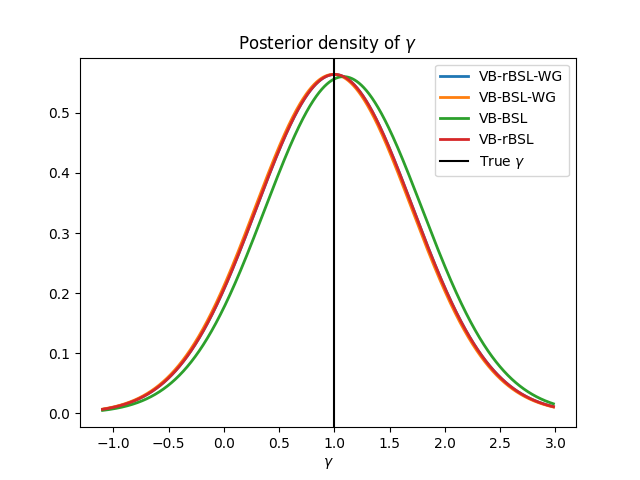}
     \end{tabular}
     \begin{tabular}{@{}c@{}}
         \centering
         \includegraphics[width=0.35\textwidth]{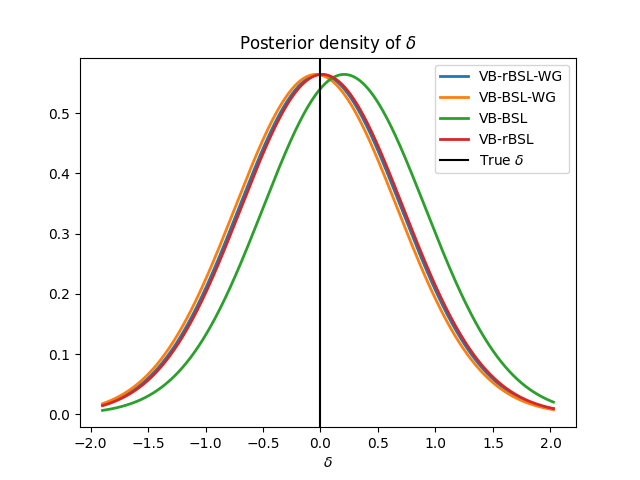}
     \end{tabular}
        \caption{$\alpha$-stable model: Marginal variational posterior distributions of the four parameters ($\alpha$, $\beta$, $\gamma$, $\delta$) for the standard BSL method and its variations (better viewed in colour).}
        \label{fig:alpha-stable-var-dist-nf-robust}
\end{figure}

Upon reviewing \autoref{tab:l2alpha-nf}, it is evident that the VB-rBSL-WG method produces the most accurate estimates. The incorporation of the WG transformation into BSL leads to a significant improvement in terms of recovering the true model parameters, particularly surpassing the performance of integrating the WG transformation in non-robust BSL method. 
\autoref{fig:alpha-stable-var-dist-nf-robust} depicts the variational distribution of the four parameters, $\alpha$, $\beta$, $\gamma$ and $\delta$. Notably, VB-BSL exhibits the poorest performance, with estimates for $\alpha$, $\beta$, and $\delta$ deviating significantly from their true values. However, other versions of VB-BSL show notable improvements in estimating all four parameters, yielding nearly identical estimates.

\subsection{g-and-k example}

The g-and-k distribution is a family of flexible distributions often used to model highly skewed data.
These distributions are defined through their quantile function \citep{rayner2002numerical}, hence their density function is not available in closed form,
but it is easy to generate data from them.
Estimating a g-and-k distribution has been a common example in the literature to test the performance of likelihood-free inference methods \citep{DROVANDI20112541}. 

The quantile function $Q(p)$, $p \in (0,1)$, of the g-and-k distribution is given as follows 
\beq\label{eq: quantile function gnk}
Q(p) = A + B \Bigr[ 1 + 0.8 \frac{1 - e^{-gz(p)}}{1 + e^{-gz(p)}} \Bigr] \big(1 + z(p)^2\big)^k z(p),
\eeq
where $z(p)$ is the quantile function of the standard normal distribution. This model has four parameters, $A$, $B$, $g$ and $k$. 
It is straightforward to generate a sample $X$ from this model using $X=Q(U)$ with $U \sim U[0,1]$. 
We generated an observed dataset of size 200 with the true parameters $\theta_\text{true}=(A, B, g, k) = (3,1,2,0.5)$. 

\begin{figure}[h!]
     \centering
     \begin{tabular}{@{}c@{}}
         \centering
         \includegraphics[width=0.45\textwidth]{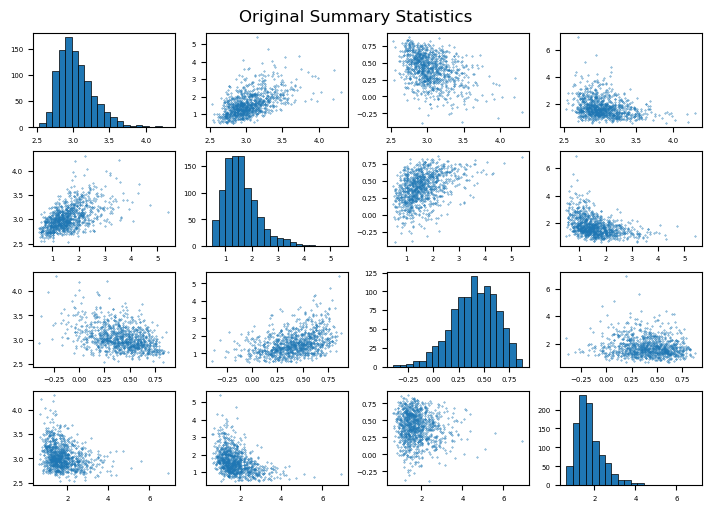}
     \end{tabular}
     \begin{tabular}{@{}c@{}}
         \centering
         \includegraphics[width=0.45\textwidth]{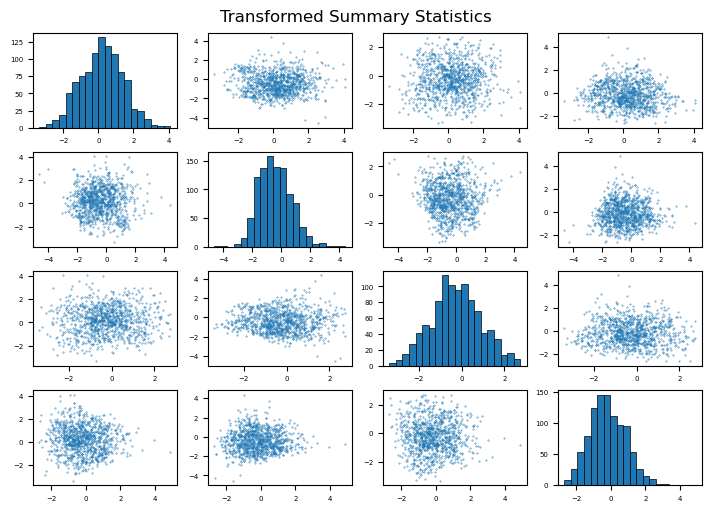}
     \end{tabular}
        \caption{g-and-k model: The left panel shows the original summary statistics. The right panel shows their WG transformation.}
        \label{fig:gnk-scatterplot-nf}
\end{figure}

For the summary statistics, we follow \cite{DROVANDI20112541} and use
\[s_A = O_4, \quad s_B = O_6 - O_2, \quad s_g = \frac{O_7 - O_5 + O_3 - O_1}{s_B}, \quad s_k = \frac{O_6 + O_2 - 2 O_4}{s_B},\]
where $O_j$ is the $j$-th octile of the data. 
We proceed to train the WG on the training set consisting of 10,000 summary statistics generated from the g-and-k distribution with $\theta_\text{true}$. The scatter plots of the validation set comprising 1,000 summary statistics, depicted in the left panel of Figure \ref{fig:gnk-scatterplot-nf}, exhibit significant deviations from multivariate normal distribution. However, the right panel illustrates the efficacy of the WG transformation, as the summary statistics now closely adhere to a normal distribution.

\begin{figure}
     \centering
     \begin{subfigure}[b]{0.35\textwidth}
         \centering
         \includegraphics[width=\textwidth]{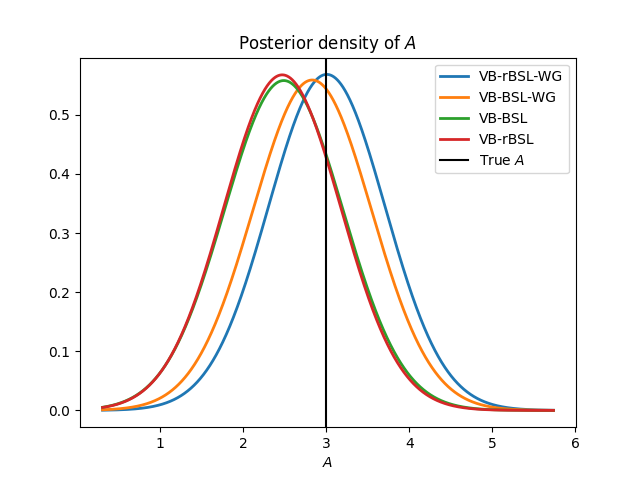}
     \end{subfigure}
     \begin{subfigure}[b]{0.35\textwidth}
         \centering
         \includegraphics[width=\textwidth]{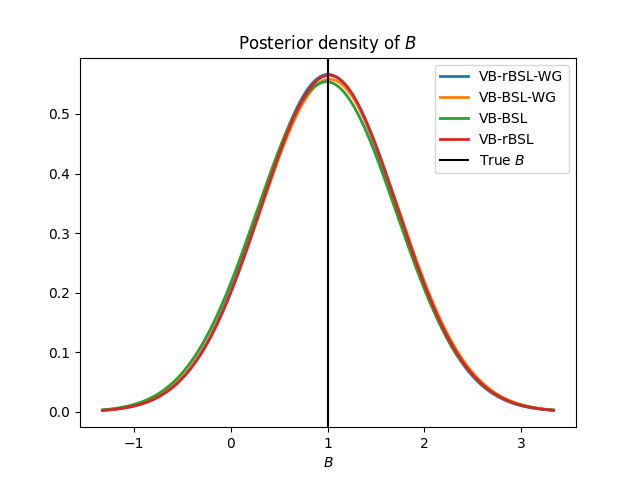}
     \end{subfigure}
     \begin{subfigure}[b]{0.35\textwidth}
         \centering
         \includegraphics[width=\textwidth]{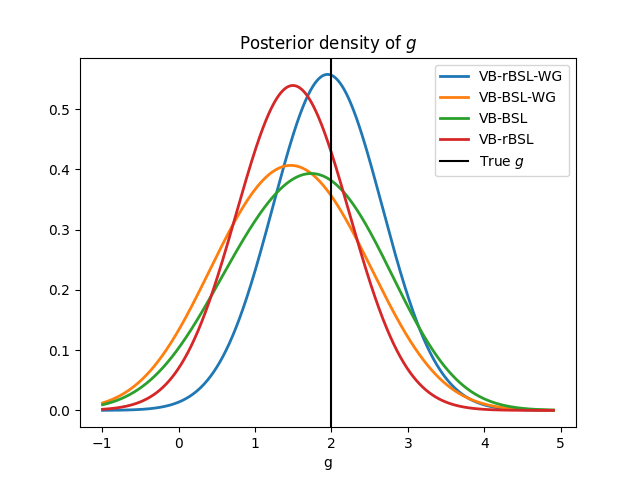}
     \end{subfigure}
     \begin{subfigure}[b]{0.35\textwidth}
         \centering
         \includegraphics[width=\textwidth]{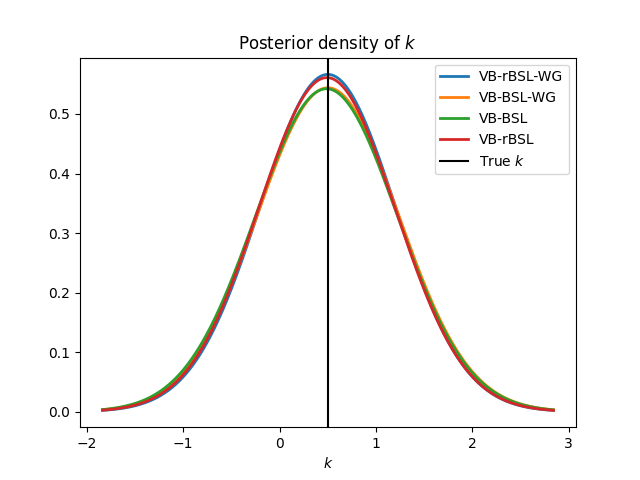}
     \end{subfigure}
        \caption{g-and-k model. Marginal variational posterior distributions for the four parameters ($A$, $B$, $g$, $k$) across the four BSL methods (better viewed in colour).}
        \label{fig:gnk-var-dist-nf-robust}
\end{figure}

Figure \ref{fig:gnk-var-dist-nf-robust} illustrates the variational distributions of the four parameters $A$, $B$, $g$, and $k$. 
In general, the plots reveal that the VB-BSL method, in comparison to the other methods, exhibits the poorest alignment of its estimates with the true model parameters. Table \ref{tab:l2norm gnk-nf} summarizes the average MSE values, across 10 different runs, between the posterior mean estimates and the true model parameters.
The results confirm the superior performance of the VB-rBSL-WG method, followed by the VB-BSL-WG and VB-rBSL.

\begin{table}[ht]
\centering
\begin{tabular}{|P{7em}|P{2.75cm}|P{2.75cm}|P{2.75cm}|P{2.75cm}|}
\hline
                                       & VB-BSL & VB-rBSL & VB-BSL-WG & VB-rBSL-WG \\ \hline
$||\theta_\text{true} - \hat{\theta}||_2$ & 0.6873 (0.0019)    & 0.7455 (0.0023)  & 0.8188 (0.0035) & 0.4211 (0.0018)    \\ \hline
MD$(||\theta_\text{true} - \hat{\theta}||)$ & 7.345 (0.0256) & 7.209 (0.0345)  & 8.943 (0.0981) & 3.956 (0.035)     \\ \hline
\end{tabular}
\caption{g-and-k example: MSE of the posterior mean estimates compared to $\theta_{\text{true}}$, over 10 different runs across the four variations of BSL methods. The numbers in brackets are standard deviations.}
\label{tab:l2norm gnk-nf}
\end{table}

\subsection{Fowler's toads example}
The study of movement patterns in amphibian animals is a topic of great interest in ecology. To investigate these patterns, \cite{marchand2017stochastic} developed a stochastic movement model based on their research of Fowler's toads at Long Point in Ontario, Canada. The model assumes that the toads hide in a refuge site during the day and forage during the night, which is a common behavior for these animals, and that every toad goes through two sequential processes. First, the overnight displacement $\Delta y$ follows a Levy-alpha stable distribution $S(\alpha, \gamma)$. Second, the toad's return behavior is modelled by a probability model 
with $p_0$ the probability that the toad selects at random one of the previous refuge sites.      
More details about the model can be found in \cite{marchand2017stochastic}. 

The radio-tracking device collects the information of $n_t$ toads on $n_d$ days to create an observation matrix $\bold{Y}$ of dimension $n_d \times n_t$. The actual data has a dimension of $(n_d = 63) \times (n_t = 66)$. 
For ease of comparison, we use a simulated data in this paper, where the observed data of size $(n_d = 63) \times (n_t = 66)$ were generated from the model at 
the true parameter $\theta_\text{true} = (\alpha, \gamma, p_0) = (1.7, 35, 0.6)$.

For the summary statistics, we follow \cite{marchand2017stochastic} and summarize the data $\bold{Y}$ into four sets $\bold{y}_1,...,\bold{y}_4$ of relative moving distances for four different time lags: 1, 2, 4 and 8 days. For example, $\bold{y}_2$ represents the displacement information within a lag of 2 days: $\bold{y}_2 =\{|\bold{Y}_{i,j}-\bold{Y}_{i+2,j}|; 1 \leq i \leq n_d - 2, 1 \leq j \leq n_t\}$. Within each set $\bold{y}_j$, we split the displacements into two subsets. The first subset consists of the displacements with absolute values less than 10 meters, in which cases the toad is considered as having returned to its original location. The number of such returns is used as the first summary statistics. For the second subset consisting of displacements with absolute values larger than 10 meters (non-returns),
we calculate the log difference between the min, median and max, leading to two summary statistics. Putting together, we obtain the summary statistics of size 3 for each lag, and hence summarizing the entire data down to 12 ($=3 \times 4$ lags) statistics.
We note that the previous literature used the log difference between eleven quantiles 0, 10th,...,100th; leading to 48 summary statistics in total.
We opt to reduce the size of the summary statistics for computational purposes.

\begin{figure}
     \centering
     \begin{tabular}{@{}c@{}}
         \centering
         \includegraphics[width=\textwidth]{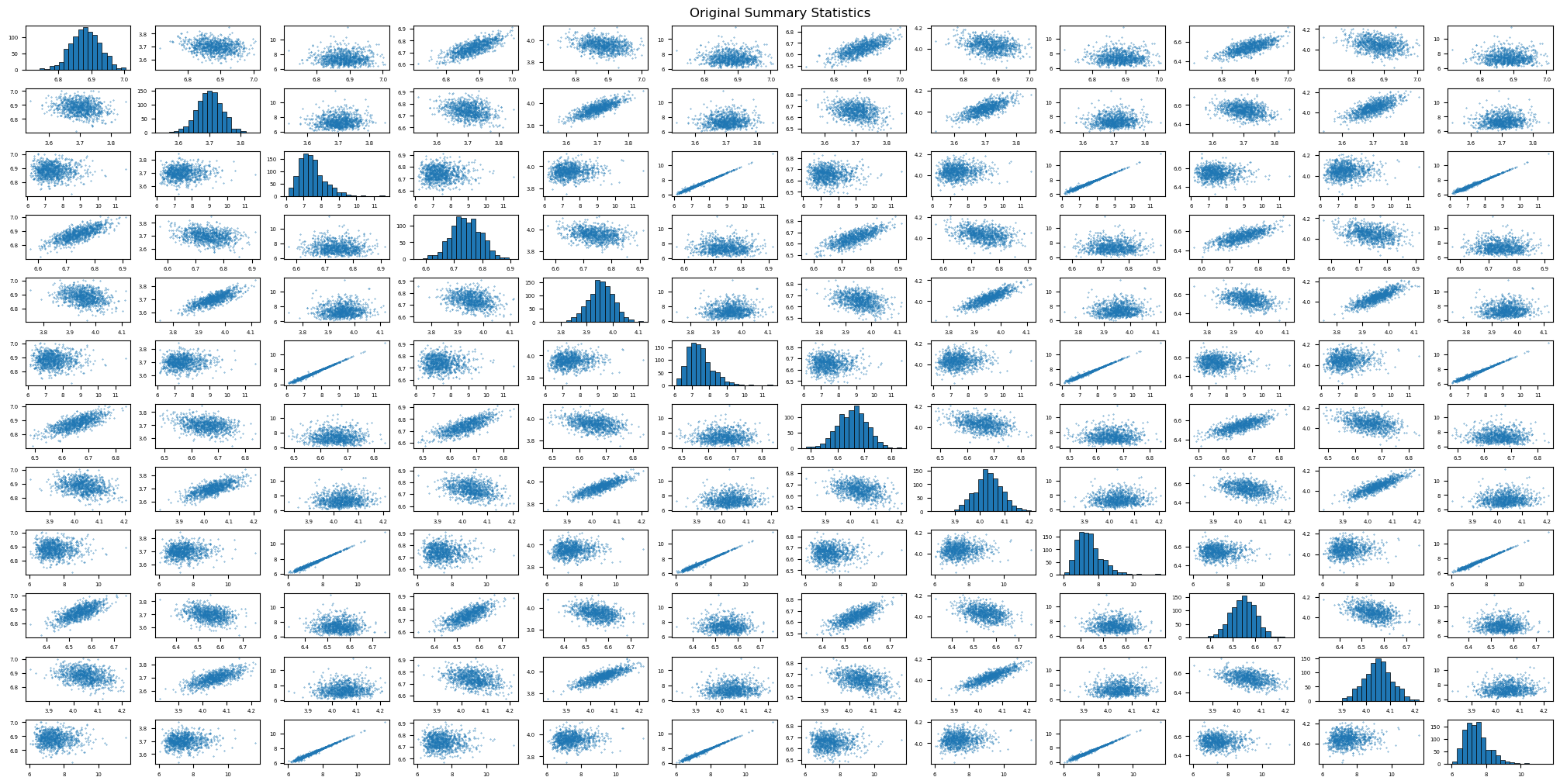}
     \end{tabular}
        \caption{Toads' movement model: scatter plots of the original summary statistics.}
        \label{fig:toad-summary-og-nf}
\end{figure}

\begin{figure}
     \centering
     \begin{tabular}{@{}c@{}}
         \centering
         \includegraphics[width=\textwidth]{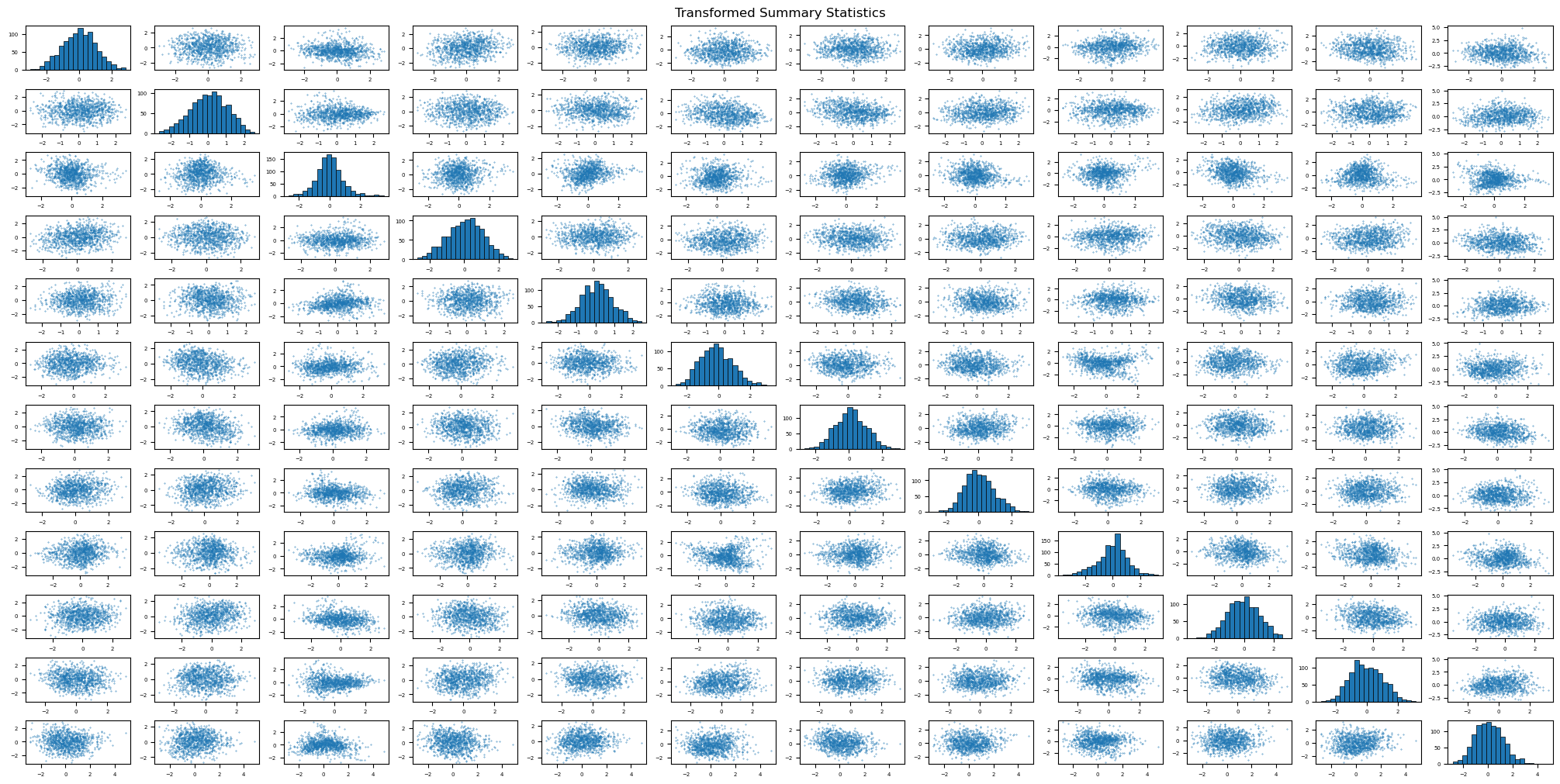}
     \end{tabular}
        \caption{Toads' movement model: scatter plots of the WG transformed summary statistics.}
        \label{fig:toad-summary-transformed-nf}
\end{figure}

We apply the resulting WG transformation to the validation set: while the original summary statistics deviate from a multivariate normal distribution as shown in \autoref{fig:toad-summary-og-nf}, the transformed summary statistics exhibit elliptical shapes and adhere to a multivariate normal distribution (\autoref{fig:toad-summary-transformed-nf}). To further validate these observations, we conduct the HZ normality test. The $p$-value for the original summary statistics is 0.0453, compared to 
0.1205 obtained from the transformed summary statistics. This reinforces the assertion that the transformed statistics adhere to a multivariate normal distribution.

\autoref{tab:toad-nf} presents the average MSE values and Mahalanobis distance across 10 runs, reflecting the disparity between the posterior mean estimate and the actual model parameters. Among the methods assessed, VB-BSL consistently exhibits the poorest performance. The results further indicate that the rBSL-WG method helps reduce significantly the estimation bias.

\begin{table}[ht]
\centering
\begin{tabular}{|P{7em}|P{2.75cm}|P{2.75cm}|P{2.75cm}|P{2.75cm}|}
\hline
                                      & VB-BSL & VB-rBSL & VB-BSL-WG & VB-rBSL-WG \\ \hline
$||\theta_{true} - \hat{\theta}||_2$ & 3.177 (0.022) & 1.558 (0.012) & 1.448 (0.005) & 3.033 (0.089)    \\ \hline
MD$(||\theta_{true} - \hat{\theta}||)$ & 30.872 (0.315) & 17.031 (0.166)  & 12.425 (0.149) & 29.586 (0.290)     \\ \hline
\end{tabular}
\caption{Toad model: The mean squared error (MSE) of the estimated posterior mean compared to the true parameter $\theta_{\text{true}}$, averaged across 10 distinct trials. Standard deviations are indicated in parentheses.}
\label{tab:toad-nf}
\end{table}

\section{Conclusion}\label{sec:conclusion}
Our WG transformation combined with rBSL and efficient VB provides a powerful and flexible approach to inference in likelihood-free problems that overcomes many of the limitations of existing methods. We hope that this paper will encourage the wider adoption of likelihood-free methods in a range of applications, where they offer an attractive alternative to traditional methods.
 While our WG transformation has shown promising results, further improvements are possible for its use in high-dimensional settings. One potential avenue for improvement is to incorporate an adaptive learning rate or natural gradient, and research in this direction is currently underway.

\section*{Acknowledgement}
The research of Nhat-Minh Nguyen was partly supported by the Australian Research Council Training Centre in Data Analytics for Resources and Environment (IC190100031).
Christopher Drovandi's research was supported by an ARC Future Fellowship (FT210100260).  Christopher Drovandi is affiliated with the Queensland University of Technology Centre for Data Science. David Nott's research was supported by the Ministry of Education, Singapore, under the Academic Research Fund Tier 2 (MOE-T2EP20123-0009).  David Nott is affiliated with the Institute of Operations Research and Analytics, National University of Singapore.

\section*{Appendix A: Differential calculus on the Wasserstein space}
\subsection*{A1. Wasserstein space} 
Recall that Wasserstein space $\mathbb{W}_2(\SetR^d)$ is the space of probability measures $\mathcal P_2^\text{ac}(\SetR^d)$, equipped with the 
Wasserstein distance 
\beq\label{eq:Wasserstein distance}
W_2(\mu,\nu)=\Big\{\inf_{T:T_{\#}\mu=\nu}\int_{\SetR^d}\|x-T(x)\|^2\mu(dx)\Big\}^{1/2},\;\;\mu,\nu\in\mathcal P_2^\text{ac}(\SetR^d).
\eeq
The push measure $\nu=T_{\#}\mu$ means $\nu(B)=\mu\big(\{x:T(x)\in B\}\big)$ for all measurable $B\subset \SetR^d$, or
\[\int \varphi(y)T_{\#}\mu(dy)=\int\varphi(T(x))\mu(dx),\;\;\forall\;\nu-\text{integrable function $\varphi$}.\]
To avoid technical complication, we limit ourselves to absolutely continuous measures in this paper.
By Brenier's theorem \citep{brenier1991polar}, 
the infimum in \eqref{eq:Wasserstein distance} is attained by an unique optimal map, denoted as $T_{\mu}^{\nu}$, which equals $\mu$-a.s. to the gradient of a convex function on $\SetR^d$.
That optimal map pushes $\mu$ to $\nu$ in the sense that $T_{\mu}^{\nu}(X)\sim \nu$ if $X\sim\mu$.
The optimal map $T_{\nu}^{\mu}$ also exists uniquely and that $T_{\mu}^{\nu}=(T_{\nu}^{\mu})^{-1}$ $\mu$-a.s. \citep{lanzetti2022first}[Proposition 2.2].

Fix a measure $\mu\in \mathcal P_2^\text{ac}(\SetR^d)$ and let $\varphi$ be a convex function on $\SetR^d$. Let $\nu=(\nabla\varphi)_{\#}\mu$.
As a consequence of Brenier's theorem, $T_{\mu}^{\nu}=\nabla\varphi$ $\mu$-a.s. That is, there might be other maps pushing $\mu$ to $\nu$, but 
only $\nabla\varphi$ is the optimal map in the sense of \eqref{eq:Wasserstein distance}.
This fact is useful in the Wasserstein calculus below.

\subsection*{A2. Differentiability} 
We wish to extend calculus concepts such as differentiability, gradient and convexity of functions defined on the Euclidean space $\SetR^d$ to those of functionals defined on the Wasserstein space $\mathbb W_2(\SetR^d)$. 

Let $F$ be a functional defined on $\mathbb W_2(\SetR^d)$. It is said to be {\it differentiable} at $\mu$ if there exists a map $\xi\in L^2(\SetR^d;\mu)$, i.e. $\int\|\xi(x)\|^2d\mu(x)<\infty$, such that for any $\nu\in\mathbb W_2(\SetR^d)$,
\beq\label{eq:differentiability}
F(\nu)-F(\mu)=\langle\xi,T_{\mu}^{\nu}-\text{Id}\rangle_\mu+o\big(W_2(\mu,\nu)\big),
\eeq
where $T_{\mu}^{\nu}$ is the optimal map from $\mu$ to $\nu$. 
Here, $\langle\cdot,\cdot\rangle_\mu$ denotes the inner-product on the vector space $L^2(\SetR^d;\mu)$, i.e.,
\[\langle\zeta,\eta\rangle_\mu=\int\langle\zeta(x),\eta(x)\rangle d\mu(x);\;\;\;\;\zeta,\eta\in L^2(\SetR^d;\mu), \]
with $\langle\cdot,\cdot\rangle$ the standard inner-product on $\SetR^d$.
The map $\xi$ in \eqref{eq:differentiability}, if exists, is unique $\mu$-a.s. and called the {\it Wasserstein gradient} of $F$ at $\mu$, denoted as $\nabla_\mu F$.

Consider a function $\phi:\SetR^d\to\SetR$ such that the eigenvalues of the Hessian $\nabla^2\phi$ are uniformly bounded by some finite constant. 
As $\nabla T=I_d+\epsilon \nabla^2\phi\geq0$ for small enough $\epsilon$,
$T=\text{Id}+\epsilon\nabla\phi$ must be the gradient of a convex function. Then $T$ is the optimal map from $\mu$ to $\nu=T_\#\mu$. From \eqref{eq:differentiability},
\beqn
\lim_{\epsilon\to0}\frac{F\big((\text{Id}+\epsilon\nabla\phi)_\#\mu\big)-F(\mu)}{\epsilon}=\langle\nabla_\mu F,\nabla\phi\rangle_\mu.
\eeqn
This therefore provides a first-order approximation of $F$
\beq\label{eq:first-order approximation}
F\big((\text{Id}+\epsilon\nabla\phi)_\#\mu\big)= F(\mu) + \epsilon\langle\nabla_\mu F,\nabla\phi\rangle_\mu + o(\epsilon).
\eeq
The vector space of such mappings $\nabla\phi$ can be thought of as the {\it tangent space} of $\mathbb W_2(\SetR^d)$ at $\mu$, denoted as $\mathcal{T}_\mu\mathbb W_2(\SetR^d)$.
Together with the inner-product $\langle\cdot,\cdot\rangle_\mu$ equipped on $\mathcal{T}_\mu\mathbb W_2(\SetR^d)$, $\mathbb W_2(\SetR^d)$ becomes a Riemannian manifold \citep{villani2009optimal}.
On this manifold, geometric objects such as geodesic curve, length, Riemannian gradient, etc. can be defined.
Note that, the Wasserstein gradient defined above coincides with the Riemannian gradient defined by viewing $\mathbb W_2(\SetR^d)$ as a Riemannian manifold.

We also can define the {\it directional derivative} of $F$ at $\mu$ along the ``direction" $\nabla\phi$ as
\beq\label{eq:Wasserstein directional derivative}
DF[\mu](\nabla\phi) = \lim_{\epsilon\to0}\frac{F\big((\text{Id}+\epsilon\nabla\phi)_\#\mu\big)-F(\mu)}{\epsilon}=\langle\nabla_\mu F,\nabla\phi\rangle_\mu.
\eeq
One can go further and obtain a second-order approximation of $F\big((\text{Id}+\epsilon\nabla\varphi)_\#\mu\big)$ as
\beq\label{eq:2nd-order approximation}
F\big((\text{Id}+\epsilon\nabla\phi)_\#\mu\big) = F(\mu)+\epsilon\langle\nabla_\mu F,\nabla\phi\rangle_\mu+\frac{\epsilon^2}{2}\langle\text{Hess}_\mu F[\nabla\phi],\nabla\phi\rangle_\mu+o(\epsilon^2)
\eeq
where $\text{Hess}_\mu F$ is the Hessian operator of $F$ at $\mu$. Equation \eqref{eq:2nd-order approximation} can be useful for second-order optimization on the Wasserstein space.

\paradot{Example: potential energy} Consider the {\it potential energy}
\[\mathcal V(\mu)=\int_{\SetR^d} V(x)\mu(dx),\]
assuming that $V(x)$ is twice differentiable and $\nabla V$ is $\beta$-Lipschitz continuous on $\SetR^d$.
It can be shown that $\nabla_\mu\mathcal V=\nabla V$ and that \citep{villani2009optimal}[Chapter 15]  
\beq
\langle\text{Hess}_\mu \mathcal V[\nabla\phi],\nabla\phi\rangle_\mu=\langle\nabla^2V[\nabla\phi],\nabla\phi\rangle_\mu
\eeq
with $\nabla^2 V$ the Euclidean Hessian of $V$. As $\nabla^2 V\leq \beta I_d$, 
\[\langle\text{Hess}_\mu \mathcal V[\nabla\phi],\nabla\phi\rangle_\mu\leq \beta\|\nabla\phi\|^2_\mu.\]
Therefore, $\mathcal V$ is $\beta$-smooth on the Wasserstein space; see \eqref{eq:Wasserstein beta-smooth} for the definition.

\paradot{Example: negative entropy} Consider the {\it negative entropy}
\beq\label{eq:negative entropy}
\mathcal H(\mu) = \int \mu(x)\log\mu(x)dx.
\eeq
It can be shown that \citep{villani2009optimal} $\nabla_\mu \mathcal H=\nabla_x\log \mu(x)$ and 
\beq\label{eq:Hessian negative entropy}
\langle\text{Hess}_\mu \mathcal H[\nabla\phi],\nabla\phi\rangle_\mu=\int\tr\big[(\nabla^2\phi)^2\big]d\mu=\int\|\nabla^2\phi\|_F^2d\mu,
\eeq
as $\nabla^2\phi$ is symmetric, $\tr\big[(\nabla^2\phi)^2\big]=\|\nabla^2\phi\|_F^2$ - the Frobenius norm of $\nabla^2\phi$.
Because of \eqref{eq:Hessian negative entropy}, it is not clear if $\mathcal H$ is smooth on the Wasserstein space; see, also, \cite{wibisono2018sampling}.
 
\subsection*{A3. Geodesic convexity} 
Let $T_{\mu}^{\nu}$ be the optimal map pushing $\mu$ to $\nu$.
For each $t\in[0,1]$, define $\mu_t=\big(\text{Id}+t(T_{\mu}^{\nu}-\text{Id})\big)_\#\mu$, then $(\mu_t)_{0\leq t\leq 1}$ is a curve 
on $\mathbb W_2(\SetR^d)$.
This is a geodesic curve connecting $\mu$ and $\nu$ \citep{villani2021topics}[Chapter 8], i.e., it has the minimum length among all the curves between $\mu$ and $\nu$. 
A functional $F:\mathbb W_2(\SetR^d)\to\SetR$ is said to be {\it $\alpha$-geodesically convex}, $\alpha\in\SetR$, if
\beq\label{eq:lambda-geodesically convex 1}
F(\mu_t)\leq (1-t)F(\mu)+tF(\nu)-\frac{\alpha}{2}t(1-t)W_2^2(\mu,\nu),\;\;t\in[0,1].
\eeq
When $\alpha=0$, $F$ is said to be geodesically convex, and strongly geodesically convex if $\alpha>0$.
If $F$ is geodesically convex, its minimizer is unique.  
If $F$ is also differentiable at $\mu$, then \eqref{eq:lambda-geodesically convex 1} implies
\beq\label{eq:lambda-geodesically convex 2}
F(\nu)\geq F(\mu)+\langle\nabla_\mu F,T_{\mu}^{\nu}-\text{Id}\rangle_\mu+\frac{\alpha}{2}W_2^2(\mu,\nu),\;\;\;\forall\mu,\nu.
\eeq

\paradot{Examples} If $V$ is $\alpha$-convex on $\SetR^d$, then $\mathcal V$ is $\alpha$-geodesically convex. 
Indeed, 
\bean
\mathcal V(\mu_t)&=&\int V\big((1-t)x+tT_{\mu}^{\nu}(x)\big)d\mu\\
&\leq&\int\Big((1-t)V(x)+tV(T_{\mu}^{\nu}(x))-\frac{\alpha}{2}t(1-t)\|x-T_{\mu}^{\nu}(x)\|^2\Big)d\mu(x)\\ 
&=&(1-t)\mathcal V(\mu)+t\mathcal V(\nu)-\frac{\alpha}{2}t(1-t)W_2^2(\mu,\nu).
\eean
It can be shown that $\mathcal H$ is geodesically convex \cite{villani2009optimal}. 

\subsection*{A4. Smoothness and Wasserstein gradient descent}
A functional $F:\mathbb W_2(\SetR^d)\to\SetR$ is said to be {\it $\beta$-smooth} if for any $\mu,\nu\in\mathbb W_2(\SetR^d)$
\beq\label{eq:Wasserstein beta-smooth}
F(\nu)\leq F(\mu)+\langle\nabla_\mu F,T_{\mu}^{\nu}-\text{Id}\rangle_\mu+\frac{\beta}{2}W_2^2(\mu,\nu).
\eeq
If the Hessian operator of $F$ is bounded by $\beta$, $F$ is $\beta$-smooth.

Just like smoothness guarantees convergence of gradient descent in $\SetR^d$, the smoothness defined in \eqref{eq:Wasserstein beta-smooth} guarantees the convergence of {\it Wasserstein gradient descent} in $W_2(\SetR^d)$. Starting from an initial measure $\mu^{(0)}$, Wasserstein gradient descent for optimizing $F$ iterates as follows:
\[\mu^{(k+1)}=\big(\text{Id}-\epsilon\nabla_{\mu^{(k)}}F\big)_\#\mu^{(k)},\;\;k=0,1,...\]
where $\epsilon>0$ is a step size, and $\nabla_\mu F$ is the Wasserstein gradient of $F$ at $\mu$.

Indeed, suppose that the $d\times d$ matrix $\nabla(\nabla_\mu F)$ has its eigenvalues uniformly bounded by $L>0$ for all $\mu$, then $T=\text{Id}-\epsilon\nabla_\mu F$ is the gradient of a convex function provided that $\epsilon<1/L$. This shows that $T_{\mu^{(k)}}^{\mu^{(k+1)}}=\text{Id}-\epsilon\nabla_{\mu^{(k)}}F$ is the optimal map from $\mu^{(k)}$ to $\mu^{(k+1)}$. Using \eqref{eq:Wasserstein beta-smooth} for $\mu=\mu^{(k)}$ and $\nu=\mu^{(k+1)}$, we have
\[F\big(\mu^{(k+1)}\big)\leq F\big(\mu^{(k)}\big)-\epsilon\|\nabla_{\mu^{(k)}}F\|_{\mu^{(k)}}+\frac{\beta\epsilon^2}{2}\|\nabla_{\mu^{(k)}}F\|_{\mu^{(k)}},\]
or
\beq\label{eq:functional reduce}
F\big(\mu^{(k+1)}\big)-F\big(\mu^{(k)}\big)\leq -\epsilon(1-\frac12\epsilon\beta)\|\nabla_{\mu^{(k)}}F\|_{\mu^{(k)}},
\eeq
which guarantees that the value of the objective functional $F$ is reduced in each iteration of Wasserstein gradient descent.
Given \eqref{eq:functional reduce}, together with the geodesical $\alpha$-convexity of $F$, convergence of Wasserstein gradient descent can be established.


\section*{Appendix B: Proofs}
\begin{proof}[Proof of Proposition \ref{lemma}]
We prove \eqref{eq:recursive} by induction. It is obviously true for $s=0$ by definition; assume that it is true for some $s>0$, i.e. $A_s=\Psi_s^{-1}$.
Using the Sherman-Morrison formula, we have that
\begin{align*}
A_{s+1}&= A_{s}-\big(1+\psi_{s+1}^\top A_{s}\psi_{s+1}\big)^{-1}A_{s}\psi_{s+1}\psi_{s+1}^\top A_{s}\\
&=\left( A_s^{-1}+\psi_{s+1}\psi_{s+1}^\top \right)^{-1}\\
&=\left(\Psi_0+\sum_{j=1}^{s+1} \psi_j\psi_j^\top\right)^{-1}\\
&=\Psi_{s+1}^{-1}.
\end{align*}
Hence, $A_s = \Psi_s^{-1}$ for all $s$.
For any $\theta$, by the law of large numbers,
\[\wh P(\theta)^{-1}=\frac1N\Big(\Psi_0+\sum_{j=1}^N(s_j-\hat\mu(\theta))(s_j-\hat\mu(\theta))^\top\Big)\xrightarrow[N \to + \infty]{a.s} \Sigma(\theta)\]
and hence, $\wh P(\theta)\xrightarrow[N \to + \infty]{a.s}\Sigma(\theta)^{-1}$.
\end{proof}

\begin{proof}[Proof of Proposition \ref{pro:JKO convergence}]
As $\mu^{(k+1)}$ is the minimizer in \eqref{eq:JKO},
\[F(\mu^{(k+1)})-F(\mu^{(k)})\leq-\frac{1}{2\epsilon}W_2^2(\mu^{(k+1)},\mu^{(k)}).\]
By the optimility condition, the Wasserstein gradient of $F(\mu)+\frac{1}{2\eps}W_2^2(\mu,\mu^{(k)})$ at $\mu=\mu^{(k+1)}$ must vanish.
That is
\[\nabla_{\mu^{(k+1)}}F+\frac{1}{\epsilon}\big(\text{Id}-T_{\mu^{(k+1)}}^{\mu^{(k)}}\big)=0\]
where $T_{\mu^{(k+1)}}^{\mu^{(k)}}$ is the optimal map pushing $\mu^{(k+1)}$ to $\mu^{(k)}$.
See, e.g., \cite{lanzetti2022first}[Corollary 2.11] for the Wasserstein gradient of the squared Wasserstein distance.
It implies
\[
\epsilon^2\E_{\mu^{(k+1)}}\|\nabla_{\mu^{(k+1)}}F\|^2 = \E_{\mu^{(k+1)}}\big\|\text{Id}-T_{\mu^{(k+1)}}^{\mu^{(k)}}\big\|^2 = W_2^2(\mu^{(k+1)},\mu^{(k)}),
\]
and hence,
\beq\label{eq: JKO resulting}
F(\mu^{(k+1)})-F(\mu^{(k)})\leq-\frac{\epsilon}{2}\E_{\mu^{(k+1)}}\|\nabla_{\mu^{(k+1)}}F\|^2.
\eeq
As $\alpha I \preccurlyeq \nabla^2V(x)$, by the Bakry-Emery theorem \citep{bakry2006diffusions,otto2000generalization} we have that, and any $\mu\in\mathcal P_2^\text{ac}(\SetR^d)$
\beq\label{eq:LSI}
F(\mu)\leq \frac{1}{2\alpha}\E_{\mu}\|\nabla_\mu F\|^2
\eeq
which is known as the $\alpha$-gradient domination, also called the logarithmic Sobolev inequality with constant $\alpha$.
Combining \eqref{eq: JKO resulting} and \eqref{eq:LSI},
\beq\label{eq:Lemma JKO}
F(\mu^{(k+1)})\leq (1+\epsilon\alpha)F(\mu^{(k)}),\;\;k=0,1,...
\eeq
or
\beq\label{eq:Lemma JKO 1}
F(\mu^{(k)})\leq (1+\epsilon\alpha)^{-k}F(\mu^{(0)}),\;\;k=0,1,...
\eeq
Now, from \cite{otto2000generalization}[Theorem 1],
\[W_2^2(\mu^{(k)},\pi)\leq\frac{2}{\alpha}F(\mu^{(k)})\leq \frac{2}{\alpha}(1+\epsilon\alpha)^{-k}F(\mu^{(0)}).\]
\end{proof}

\bibliographystyle{apalike}
\bibliography{references_robust_vbsl}

\end{document}